\documentclass[12pt]{article}
\usepackage{epsf}

\usepackage{epsfig,graphics}
\usepackage {graphicx}
\usepackage {epsfig}
\usepackage {psfig}
\usepackage {subfigure}
\usepackage {tabularx} 
\usepackage{rotate}	
\usepackage{slashed}
\usepackage{cite}

\newcommand{\bmat}{\left(\begin{array}}
\newcommand{\emat}{\end{array}\right)}

\def\yzero{\smash{\hbox{$y\kern-4pt\raise1pt\hbox{${}^\circ$}$}}}
\def\p{\partial}
\def\a{\alpha}

\def\g{\gamma}

\def\beq{\begin{equation}}
\def\eeq{\end{equation}}
\def\beqa{\begin{eqnarray}}
\def\eeqa{\end{eqnarray}}

\def\-{\hphantom{-}}

\def\s2{\frac{1}{\sqrt2}}

\def\beq{\begin{equation}}
\def\eeq{\end{equation}}
\def\beqa{\begin{eqnarray}}
\def\eeqa{\end{eqnarray}}

\def\IF{\relax{\rm I\kern-.18em F}}
\def\II{\relax{\rm I\kern-.18em I}}
\def\IP{\relax{\rm I\kern-.18em P}}
\def\IC{\relax\hbox{\kern.25em$\inbar\kern-.3em{\rm C}$}}
\def\IR{\relax{\rm I\kern-.18em R}}

\def\cn{{\cal N}}

\def\Dsl{\,\raise.15ex\hbox{/}\mkern-13.5mu D} 
\def\IZ{Z\kern-.4em  Z}

\def\lam{\lambda}
\def\raw{\rightarrow}

\catcode`\@=11   
\newdimen\@rotdimen
\newbox\@rotbox  

\def\@vspec#1{\special{ps:#1}}
\def\@rotstart#1{\@vspec{gsave currentpoint currentpoint translate
   #1 neg exch neg exch translate}}
\def\@rotfinish{\@vspec{currentpoint grestore moveto}}
\def\@rotr#1{\@rotdimen=\ht#1\advance\@rotdimen by\dp#1%
   \hbox to\@rotdimen{\hskip\ht#1\vbox to\wd#1{\@rotstart{90 rotate}%
   \box#1\vss}\hss}\@rotfinish}
\def\@rotl#1{\@rotdimen=\ht#1\advance\@rotdimen by\dp#1%
   \hbox to\@rotdimen{\vbox to\wd#1{\vskip\wd#1\@rotstart{270 rotate}%
   \box#1\vss}\hss}\@rotfinish}%
\def\@rotu#1{\@rotdimen=\ht#1\advance\@rotdimen by\dp#1%
   \hbox to\wd#1{\hskip\wd#1\vbox to\@rotdimen{\vskip\@rotdimen
   \@rotstart{-1 dup scale}\box#1\vss}\hss}\@rotfinish}%
\def\@rotf#1{\hbox to\wd#1{\hskip\wd#1\@rotstart{-1 1 scale}%
   \box#1\hss}\@rotfinish}%
\def\rotate{\@ifnextchar[{\@rotate}{\@rotate[l]}}
\def\@rotate[#1]#2{\setbox\@rotbox=\hbox{#2}\@nameuse{@rot#1}\@rotbox}

\catcode`\@=12

\topmargin
-1.5cm
\textwidth
15.5cm
\textheight
23.5cm
\oddsidemargin
0.7cm
\evensidemargin
0.7cm

\begin{document}

\makeatletter
\@addtoreset{equation}{section}
\makeatother
\renewcommand{\theequation}{\thesection.\arabic{equation}}
\pagestyle{empty}
\rightline{ IFT-UAM/CSIC-12-55}
\vspace{1.5cm}
\begin{center}


\LARGE{The Intermediate Scale MSSM,  the Higgs Mass and F-theory Unification} 
  \\[3mm]
  
  \vspace{1.0cm}
  
 \large{ L.E. Ib\'a\~nez$^{a,b}$, F. Marchesano$^b$, D. Regalado$^b$ and I. Valenzuela$^{a,b}$\\[3mm]}
\small{
 $^a$Departamento de F\'{\i}sica Te\'orica  
and $^b$Instituto de F\'{\i}sica Te\'orica  UAM-CSIC,\\[-0.3em]
Universidad Aut\'onoma de Madrid,
Cantoblanco, 28049 Madrid, Spain 
\\[6mm]}
\small{\bf Abstract} \\[7mm]
\end{center}
\begin{center}
\begin{minipage}[h]{15.0cm}
Even if  SUSY   is not present at the Electro-Weak scale,  
string theory suggests its
presence at some scale $M_{SS}$ below the string scale $M_s$ to guarantee the absence of tachyons.
We explore the possible value of $M_{SS}$ consistent with gauge coupling unification and known sources
of SUSY breaking in string theory. Within F-theory $SU(5)$ unification these two requirements 
 fix  $M_{SS}\simeq 5\times 10^{10}$ GeV  at an intermediate scale and a unification scale
$M_{c}\simeq 3\times 10^{14}$ GeV.  As a direct consequence one also predicts the vanishing of the quartic Higgs
SM self-coupling at $M_{SS}\simeq 10^{11}$ GeV. This is tantalizingly consistent with recent LHC  hints of a Higgs mass
in the region 124-126 GeV. 
With such a low unification scale $M_c\simeq 3\times 10^{14}$ GeV one may worry about too fast proton decay via
dimension 6 operators. However in the F-theory GUT context $SU(5)$ is broken to the SM via hypercharge flux.
We show that this hypercharge flux deforms 
 the SM fermion wave functions leading to a suppression, avoiding in this way the strong experimental proton decay 
 constraints. In these constructions 
 there is generically an axion  with 
 a scale of size $f_a\simeq M_c/(4\pi)^2\simeq   10^{12}$ GeV which could solve the strong CP problem and provide for the observed dark matter.
 The price to pay for  these attractive features is to assume that the hierarchy problem is solved due to anthropic selection 
 in a string landscape.

\end{minipage}
\end{center}
\newpage
\setcounter{page}{1}
\pagestyle{plain}
\renewcommand{\thefootnote}{\arabic{footnote}}
\setcounter{footnote}{0}


\tableofcontents

\section{Introduction}

The LHC is already providing us with  very important information on the physics underlying
the Standard Model (SM) symmetry breaking process.  A first piece of information are the constraints on the mass
of the Higgs particle which is either heavier than 600 GeV or else confined to a region 
in the area $120-127$ GeV. In fact  after the 7 TeV run there are important  hints suggesting a Higgs mass
in the region $124-126$ GeV from both CMS and ATLAS \cite{CMSHiggs,ATLASHiggs}.
A second piece of information is the absence of
any trace of physics  beyond the Standard Model (BSM). In particular there is at present no sign of squark and gluinos 
below 1 TeV, at least if they decay via the standard  R-parity conserving channels in the Minimal
Supersymmetric Standard Model (MSSM).

A Higgs mass around $125$ GeV is in principle good news for supersymmetry.
Indeed such a value is consistent with the MSSM which predicts  a mass $<130$ GeV for its lightest 
Higgs scalar. On the other hand getting  such a Higgs mass within the MSSM typically requires a very massive 
SUSY spectrum with e.g. squarks at least  in the $3-10$ TeV region
\cite{SUSYpapers125}. This massive spectrum requires in turn a fine-tuning of
the parameters at the per-mil level. If after the run at 14 TeV the LHC sees no sign of supersymmetry or any other new physics BSM,  
the required fine-tuning will increase further and 
we will have to consider seriously the possibility that indeed the Electro-Weak (EW) scale is fine-tuned and selected 
on anthropic grounds \cite{antronucl,splitsusy,Hall:2009nd,otheranthropic}.

If the EW scale is fine-tuned and low-energy SUSY does not play a role in the hierarchy issue,  one may think 
of  resurrecting good old non-SUSY unified theories like $SU(5)$. We have to recall however the limitations 
of non-SUSY unification. Unification of gauge coupling constants,  which works so well with the MSSM, 
fails in the non-SUSY case. Furthermore in minimal $SU(5)$ models the unification scale is around $10^{14}-10^{15}$ GeV
and the proton decays too fast via dimension 6 operators. We also loose the existence of a natural candidate 
for dark matter to replace the neutralinos in the MSSM.

There is however another hint telling us that  a non-SUSY desert up to a unified $SU(5)$ scale and a fine-tuning 
of the Higgs mass  is unlikely. Indeed if a SM  Higgs mass is around $125$ GeV one knows what is the value of the 
Higgs quartic coupling $\lambda(M_{EW})$ at the EW scale. One can then extrapolate its value up in energies
using the Renormalization Group Equations (RGE). Due to the large top quark Yukawa coupling $\lambda$ decreases
at higher energies and in fact seems to vanish at a scale around $10^9-10^{12}$ GeV, with the precise 
scale depending on the precise value of the top-quark mass  $m_t$ and the strong coupling constant $\alpha_3$
 \cite{Cabrera:2011bi,Giudice:2011cg,Arbey:2011ab,EliasMiro:2011aa,Holthausen:2011aa,Wetterich:2011aa,Degrassi:2012ry,Chetyrkin:2012rz,Bezrukov:2012sa}.
If this is the case the theory becomes metastable before reaching the unification scale.

In any event,  supersymmetric or not, one expects any unified theory to be combined with gravitation into
an ultraviolet  complete theory. At present the best candidate for such completion is provided by string theory.
 It is thus natural to try to address the
unification issue within the context of string theory. In the last few years an interesting embedding of the $SU(5)$ 
unification idea into F-theory has been the subject of much work (for reviews see \cite{BOOK,ftheoryGUTs}).
These so called F-theory GUTs have some
similarities with standard $SU(5)$ field theory models but differ in some important aspects. Thus, e.g., 
the breaking of $SU(5)$ down to the SM is produced by the presence of hypercharge fluxes in the
compact dimensions instead of an explicit Higgs mechanism. 
As we will see this leads to several physical effects on both the gauge and Yukawa couplings which
modify several aspects of field theory GUTs.

In the present paper we address the embedding of the SM or its SUSY version   into the scheme of $SU(5)$ F-theory unification
without any prejudice about the size of the SUSY breaking scale $M_{SS}$.  At this scale it is assumed that soft terms
break the  SUSY SM \footnote{The SUSY spectrum could be that of the MSSM of some extension with e.g. extra 
Higgs doublets or triplets, see below. Whenever we write MSSM we also mean  this kind of extensions, which
will require minimal modifications.}
 into the minimal SM. 
We study what  the scale of unification $M_c$ and SUSY breaking $M_{SS}$ should be in order
to  obtain 1) correct gauge coupling unification, 2) sufficiently suppressed proton decay and 3) consistency
with a SM Higgs in the $124-126$ GeV region.

 F-theory unification 
has  specific hypercharge flux threshold corrections \cite{Donagi:2008ca,Blumenhagen:2008aw,Conlon:2009qa}
to the inverse couplings $\alpha_i(M_c)^{-1}$ which
are proportional to the inverse string constant $g_s^{-1}$. At strong coupling  they are suppressed
and correct gauge coupling unification is obtained with the MSSM spectrum and $M_{SS}=1$ TeV in the usual way. 
On the other hand, if one wants to remain at weak coupling,   the threshold corrections are too large and spoil MSSM
unification (unless extra effects from particles beyond the MSSM are included). However leaving the SUSY breaking
scale free with $M_{SS}\gg 1$ TeV one finds that the threshold corrections
have  the required form and size to yield correct gauge coupling unification
without the addition of any extra particle beyond the MSSM content.  
We argue that there are three  independent arguments 
suggesting $M_{SS}$ is at an intermediate scale 
$M_{SS}\simeq  5\times 10^{10}$ GeV with gauge unification then fixing a unification scale
$M_c\simeq 3\times 10^{14}$ GeV.  First, 
SUSY breaking induced by closed string fluxes gives rise generically to soft terms of order
$M_{SS}\simeq M_c^2/(\alpha_G^{1/2}M_p)\simeq 5\times 10^{10}$ GeV. Second, if indeed the SM Higgs self-coupling
vanishes at a scale of order $10^{11}$ GeV, as seems to be implied by a 
$124-126$ GeV Higgs, this may be an indication of a SUSY boundary condition 
$\lambda(M_{SS})=\frac {1}{8}(g_1^2+g_2^2)cos^22\beta$ with tan$\beta (M_{SS})  \simeq 1$.
We show that this boundary condition is quite generic in string constructions in which a Higgs
field  $H_{SM}=sin\beta H_u-cos\beta H_d^*$ is fine-tuned to be massless. Finally, in  F-theory $SU(5)$ GUT's 
there is a natural candidate for a string axion with decay constant $f_a\simeq M_c/(16\pi^2)$,
which is in the right region $\simeq 10^{12}$ GeV for axionic dark matter if $M_c\simeq 3\times 10^{14}$ GeV.

In this ISSB\footnote{Intermediate Scale Supersymmetry  Breaking.} framework
 the unification scale is relatively low, $M_c\simeq 3\times 10^{14}$ GeV and one may worry about 
fast proton decay via dimension 6 operators. Again,  the fact that the breaking of the $SU(5)$ symmetry down to the SM
proceeds due to a  hypercharge flux background rather than a  Higgs mechanism modifies the expectations. 
Indeed, in field theory the gauge coupling of the $SU(5)$ $X,Y$ gauge bosons to fermions remains unchanged before
and after symmetry breaking. However if symmetry breaking is induced by hypercharge fluxes, the $X,Y$ coupling to fermions
may be substantially suppressed due to the fact that the profile of the corresponding wave functions is modified.
We describe this novel  effect in detail by using local F-theory wave-functions of $SU(5)$ matter fields. 

The outline of this paper is as follows. In the next chapter we present a  brief review of F-theory $SU(5)$ unification
with emphasis on the particular aspects which play a role in the following sections. In chapter 3 we discuss gauge 
coupling unification and the scales  of SUSY breaking $M_{SS}$ and unification $M_c$ naturally arising. 
In chapter  4 we discuss the value of the quartic Higgs coupling at $M_{SS}$ within the context of string compactifications,
describing how tan$\beta \simeq 1$ comes naturally. In the following section we discuss how the fine-tuning of a
massless SM Higgs can proceed and the one-loop stability of the  tan$\beta \simeq 1$ boundary condition. 
In chapter 6 we address the issue of proton decay suppression and in chapter 7 we discuss other phenomenological consequences
and in particular how an axion with an appropriate  decay constant  naturally appears within this framework.
We briefly discuss the case of Split-SUSY in section 8 and  leave section 9 for some final remarks and conclusions. Three appendices complement the main text.

\section{$SU(5)$ and F-theory unification}

F-theory \cite{ftheory}  may be considered as a non-perturbative extension of Type IIB orientifold compactifications 
with 7-branes. This class of compactifications have two main phenomenological 
virtues compared to other string constructions
\cite{BOOK,ftheoryGUTs}).
First, in Type IIB 
compactifications it is well understood how moduli could be fixed in the presence of closed string
fluxes and non-perturbative effects. Secondly, particularly within F-theory, GUT symmetries like $SU(5)$ 
appear allowing for a correct structure of fermion masses (in particular a sizeable top quark mass).  Here we just review a few concepts which
are required for the understanding of the forthcoming sections. Our general discussion applies both to
perturbative Type IIB and their F-theory extensions but we will refer to them as F-theory constructions
for generality.

\begin{figure}[t]
\begin{center}
\includegraphics[width=11cm]{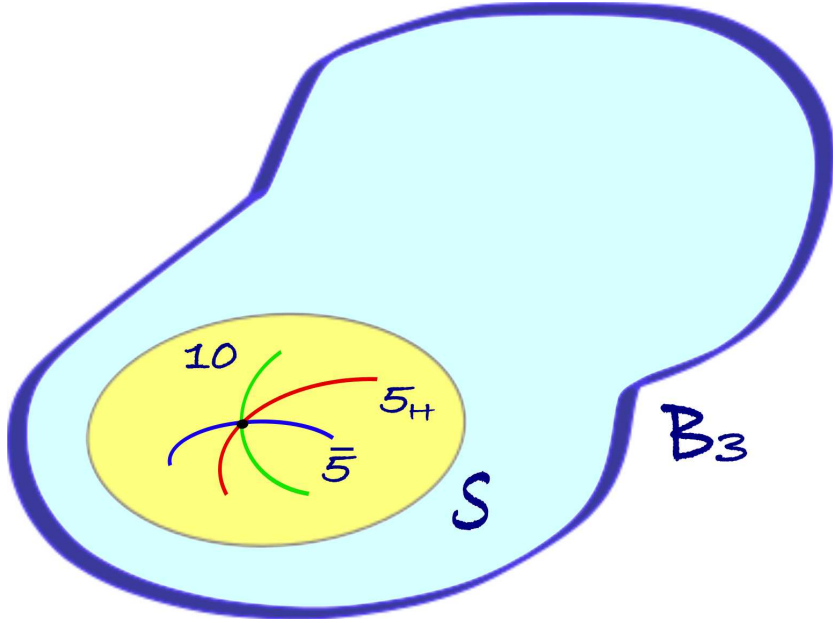}
\end{center}
\caption{Scheme of an F-theory $SU(5)$ GUT.  The six extra dimensions are compactified on $B_3$ whereas the 
$SU(5)$ degrees of freedom are localized on a  4-cycle submanifold $S$. The gauge bosons live 
on the bulk of $S$ but the chiral multiplets are localized on complex matter curves. At the intersection of
two matter curves with a Higgs curve a Yukawa coupling develops.}
\label{fthscheme}
\end{figure}
%

In Type IIB orientifold/F-theory  unified models the $SU(5)$ symmetry arises from the worldvolume fields of five 7-branes with 
their extra 4 dimensions wrapping a 4-cycle $S$  inside a six dimensional compact manifold $B_3$,  see fig.\ref{fthscheme}.
The matter fields transforming in 10-plets and 5-plets have their wave functions in extra dimensions localized
on  complex curves, the so called matter curves. These matter curves, which have two real dimensions, may be
understood as intersections  of the $SU(5)$  7-branes with extra $U(1)$ 7-branes wrapping other 4-cycles in 
$B_3$.  

In order to get an idea of the relevant mass scales one can use results from perturbative Type IIB orientifolds.
The string scale $M_s=\alpha'^{-1/2}$ is related to the  Planck scale $M_p$ by \cite{BOOK}
\beq
M_p^2\ =\ \frac {8M_s^8V_6}{(2\pi)^6g_s^2}
\label{masaplanck}
\eeq
where $V_6$ is the volume of the 6-manifold $B_3$ and $g_s$ in the string coupling constant.
$\alpha'$ is the inverse string tension.
Note that one can lower $M_s$ by having a large volume $V_6$  (or decreasing $g_s$), so that the string
scale is in principle a free parameter.

Now, the volume  $V_4$ of the 4-fold $S$ which is wrapped by the 7-branes is independent from the overall volume of $B_3$.
This volume however is related to the inverse GUT coupling constant $\alpha_G$. In particular
one has  at tree level
\beq
\frac {1}{\alpha_G} \ =\  4\pi Ref_{SU(5)}\  = \  \frac {1}{8\pi ^4g_s}\left(\frac {V_4}{{\alpha'}^2}\right)
\label{capli}
\eeq
with $f_{SU(5)}$ the gauge kinetic function. 
Parametrizing  $V_4=(2\pi R_c)^{4}$
one then has 
\beq
M_c\ =\ M_s \left(\frac {\alpha_G}{2g_s}\right)^{1/4}
\label{compactmas}
\eeq
where we define $M_c=1/R_c$.  This is slightly below the string scale (i.e. for $g_s=1/2$ and
$\alpha_G=1/38$ one has $M_c\simeq 0.4 M_s$). 
This scale $M_c$ can be identified with the GUT scale at which $SU(5)$ is broken down to the SM.
Indeed in F-theory GUTs there are no adjoint Higgs multiplets nor discrete Wilson lines available and it
is a  hypercharge flux background $<F_Y>\not=0$ which does the job \cite{BHV,DW}.
 These fluxes go through holomorphic curves $\Sigma$  inside $S$ and they are quantized, $\int_{\Sigma}F_Y=$ integer. 
Thus on dimensional grounds one has $<F_Y>\simeq 1/R_c^2=M_c^2$ and indeed one can identify $M_c$ with the GUT scale.\footnote{Hypercharge fluxes have an additional use in F-theory GUT's. Indeed, by appropriately choosing these open string  fluxes 
one can get doublet-triplet splitting in the $SU(5)$ Higgs 5-plet, see refs. \cite{BHV,DW,ftheoryGUTs,BOOK} for details.
However, as we will remark later, doublet-triplet splitting is not strictly necessary in our scheme, and so the constraints
that are usually required on the hypercharge flux in order to achieve doublet-triplet splitting can be relaxed in our setup.}

We want to consider here  the case in which slightly below the unification scale we have
unbroken $N=1$ SUSY with an MSSM spectrum (or some slight generalization, see below). One reason for that assumption is
that such class of vacua have no tachyons which could cause any premature 
instability in the theory.  We will find additional a posteriori justifications for such an
option in the forthcoming chapters. We will however allow for  SUSY to be broken in the
MSSM sector at a scale $M_{SS}$ to be determined. It is however important to realize that there
is a natural scale of SUSY breaking in Type IIB/F-theory compactifications.

Indeed, a most natural source of SUSY breaking  is the presence of closed string fluxes in such vacua.
More precisely, it is well known that e.g. generic  RR and NS 3-form fluxes  $G_3$ in Type IIB orientifolds induce SUSY-breaking soft terms
\cite{softfromflux}.
These are also the fluxes which play a prominent role in the fixing of the moduli in these vacua.
Since these fluxes  live in the full CY and are quantized on 3-cycles, the said soft terms scale
like $G_3\simeq c\, \alpha ' /(V_6^{1/2})$. One thus  finds for the size of soft terms \cite{BOOK}
\beq
M_{SS}\simeq  \frac {g_s^{1/2}}{\sqrt{2}}G_3 \ =\ 
\frac {cg_s^{1/2}}{\sqrt{2}}\frac {\alpha '}{V_6^{1/2}}\ =\ \frac {cM_s^2}{M_p}
\eeq
with c some fudge factor. 
Taking $c=1$ and taking into account eq.(\ref{compactmas}) one thus has an estimation for $M_{SS}$
\footnote{Note that setting $M_s\simeq 10^{11}$ GeV one would obtain a scheme 
with soft terms around 1 TeV, which would be consistent with a SUSY solution of the hierarchy problem.
However that would require also a string scale of order $M_s\simeq 10^{11}$ GeV and MSSM gauge coupling unification
would be lost. Alternatively one can set $M_c\simeq M_s\simeq 10^{16}$ GeV consistent with MSSM gauge coupling unification if 
the effect of fluxes is somehow suppressed. Possible ways to suppress it would be assuming some fine-tuning 
in the flux  or some warp factor leading to a flux dilution. This is the implicit assumption in models with flux induced 
SUSY breaking, $M_s\simeq 10^{16}$ GeV and a standard SUSY solution to the hierarchy problem.}
\beq
M_{SS}\ = \  \frac {(2g_s)^{1/2}}{\alpha_G^{1/2}} \frac {M_c^2}{M_p} \ .
\label{genericmss}
\eeq
We will discuss other possible sources of SUSY breaking in section 4.

Although it will not play a relevant role in our discussion, let us briefly mention how the Yukawa couplings 
appear in the framework of F-theory unification. As we said the matter multiplets of the MSSM are confined
in complex matter curves within the 4-fold $S$. Yukawa couplings appear at triple intersection points in $S$ 
in which two matter curves involving 10-plets and  5-plets cross with a matter curve containing the
Higgs 5-plets, see fig.\ref{fthscheme}. The Yukawa
 couplings may be computed as in standard Kaluza-Klein compactifications from triple overlap integrals of the form
 \beq
 {\cal Y}_{D,L}^{ij} \ =\ \int_S \Psi_{10}^i\Psi_{\bar 5}^j\Phi_{H_D} \quad  \quad \quad  {\cal Y}_{U}^{ij} \ =\ \int_S \Psi_{10}^i\Psi_{10}^j\Phi_{H_U}  \ .
 \label{yukawas}
\eeq
where $i,j$ are family indices. The wave functions have a Gaussian profile so that one only needs local information 
about these wave functions  around the intersection points  in order to compute the Yukawa couplings.
This local information may be extracted from the local   equations of motion which may be solved so that
quite explicit expressions for these wave functions may be obtained. 
We will use these local wave functions 
when discussing the proton decay suppression in section 6.

As a summary we see that F-theory $SU(5)$ unification allows for a general structure of scales 
$M_{SS}< M_c < M_s<M_p$. In what follows we will discuss how constraints from gauge coupling unification
and the Higgs mass fix these scales.

\section{Gauge coupling unification and hypercharge flux}

In order to check for gauge coupling unification we will 
 assume that at some scale $M_c$ the 
$SU(5)$ symmetry is broken by hypercharge fluxes down to the SM group.   We will asume that after this breaking the
particle content is that of the MSSM (although we will allow for some variation below). However, unlike in the usual low energy SUSY scenario, we allow the scale
of SUSY breaking on the MSSM $M_{SS}$ to be a free parameter. 
We know that the standard MSSM prediction  for gauge coupling unification \cite{SUSYGUTs}
 with
$M_{SS}\simeq 1$ TeV is quite successful. 
On the other hand for $M_{SS}\simeq M_c$ we have the 
SM below $M_c$ and we know that coupling unification fails. On the basis of this one could 
conclude that gauge coupling unification forces $M_{SS}$ to be close to the weak scale.
Interestingly enough,  
the breaking of the $SU(5)$ symmetry via  fluxes has a novel type of threshold corrections 
\cite{Donagi:2008ca,Blumenhagen:2008aw,Conlon:2009qa}
compared to the field theory case, as we now describe.\footnote{These corrections result from the expansion in 
powers of fluxes $F$ of the  Dirac-Born-Infeld 
plus Chern-Simmons (DBI+CS) action of the 7-branes, see e.g.\cite{BOOK} for a review.}
To leading order the gauge kinetic function for the $SU(5)$ group within the  7-branes is given by
the local K\"ahler modulus $T$ whose real part is proportional to $V_4$, consistently 
with eq.(\ref{capli}). However in the presence of hypercharge fluxes $F_Y$ 
the gauge kinetic functions get corrections \cite{Blumenhagen:2008aw}
 \beqa
4\pi f_{SU(3)}\ &=& T\ -\frac{1}{2}\tau \int_S{ F}_a\wedge {F}_a \label{lastres}\\
4\pi  f_{SU(2)}\ & = & T - \frac{1}{2} \tau \int_S \left({ F}_a\wedge { F}_a+{ F}_Y\wedge{F}_Y\right) \nonumber\\
\frac {3}{5} 4\pi f_{U(1)}\ & = & T - \frac{1}{2} \tau\int_S \left({ F}_a\wedge { F}_a+\frac {3}{5}({ F}_Y\wedge {F}_Y)\right) \ .
\label{3corr}
\nonumber
\eeqa
where $\tau =\frac {1}{g_s}+iC_0$ is the complex dilaton and  ${ F}_a$ are fluxes along the U(1) contained in
the U(5) gauge group of the 7-branes which are needed for technical reasons but are not relevant in our discussion. 
It turns out that in order to get rid of exotic matter massless fields beyond those of the  MSSM 
the topological condition $\int  F_Y\wedge {F}_Y=-2$ should be fulfilled \cite{BHV,DW}. This implies that at the compactification scale 
one has the condition
\beq
\frac {1}{\alpha_1(M_c)} \ = \ \frac {1}{\alpha_2(M_c)} \  + \  \frac {2}{3\alpha_3(M_c)} .
\label{couplillas}
\eeq
which is a generalization of the standard relationship $5/3\alpha_1=\alpha_2=\alpha_3$. 
In addition one also obtains
\beq
\frac{3}{5}\frac{1}{g_s}\ =\      \frac{3}{5\alpha_1(M_c)}\ -\ \frac{1}{\alpha_3(M_c)} \ =\ \frac {3}{5} \left( \frac{1}{\alpha_2(M_c)}\ -\ \frac{1}{\alpha_3(M_c)}\right) \ .
\label{calculogs}
\eeq
Thus the size of the threshold corrections is determined by the inverse of the
string coupling $g_s$. The  corrections by themselves would imply 
an ordering of the size of the fine structure constants
at  $M_c$  given by
\beq
\frac {1}{\alpha_3(M_c)} \ < \frac {1}{\alpha_1(M_c)} \  < \frac {1}{\alpha_2(M_c)} \ .
\eeq
We will be neglecting in what follows other possible sources of threshold corrections like 
the loop contributions of KK massive modes.
The corrections in eq.(\ref{3corr})  may in fact spoil the standard joining of gauge coupling constants in the MSSM, which works quite well, 
unless they are suppressed  by assuming  $g_s\gg 1$.
 Furthermore the above ordering of couplings 
goes in the wrong direction if one wanted to use such corrections to further improve the agreement
with experiment \cite{Blumenhagen:2008aw}.

Interestingly enough,  in our setting with undetermined $M_{SS}$ the corrections have just the required 
form and size to get consistency with gauge coupling unification without the addition of any extra matter field
beyond the MSSM (see also ref.\cite{Huo:2010fi}).
The one-loop renormalization group equations lead to the 
standard formulae
\beq
\frac {1}{\alpha_i(M_c)} \ =\ 
\frac {1}{\alpha_i(M_{EW})}       \ -\ 
\frac {b_i^{NS}}{2\pi }\ log\frac {M_{SS}}{M_{EW}} \ -\ 
\frac {b_i^{SS}}{2\pi }\ log\frac {M_{c}}{M_{SS}}
\label{betas}
\eeq
where $b_i^{NS},b_i^{SS}$ are the one-loop beta-function coefficients of the SM and the MSSM respectively.
These are given by $(b_1,b_2,b_3)^{NS}=(41/6,-19/6,-7)$  and $(b_1,b_2,b_3)^{SS}=(11,1,-3)$.
Combining these equations and including the boundary condition (\ref{couplillas}) (which amounts to allowing
for the above threshold corrections) one obtains 
\beqa
2\pi \  \left(\frac {1}{\alpha_1(M_{EW})} 
- \frac {1}{\alpha_2(M_{EW})}  -\frac {2}{3\alpha_3(M_{EW})} \right)   \  &= & \\  \nonumber
\ =\ \left(b_1^{NS}-b_2^{NS}-\frac {2}{3}b_3^{NS} \right)log \left(\frac {M_{SS}}{M_{EW}}\right) 
&+ & \left(b_1^{SS}-b_2^{SS}-\frac {2}{3}b_3^{SS}\right)log\left(\frac {M_{c}}{M_{SS}}\right) 
\label{rgeunif}
\eeqa
In our case this yields
\beq
\frac {44}{3}\ log \frac {M_{SS}}{M_{EW}} \ +\ 12 \  log \frac {M_{c}}{M_{SS}} \ =\ 2\pi \  \left(\frac {1}{\alpha_1(M_{EW})} 
- \frac {1}{\alpha_2(M_{EW})}  - \frac {2}{3\alpha_3(M_{EW})} \right) .
\label{lineaunif}
\eeq
%
\begin{figure}[t]
\begin{center}
\includegraphics[width=13cm]{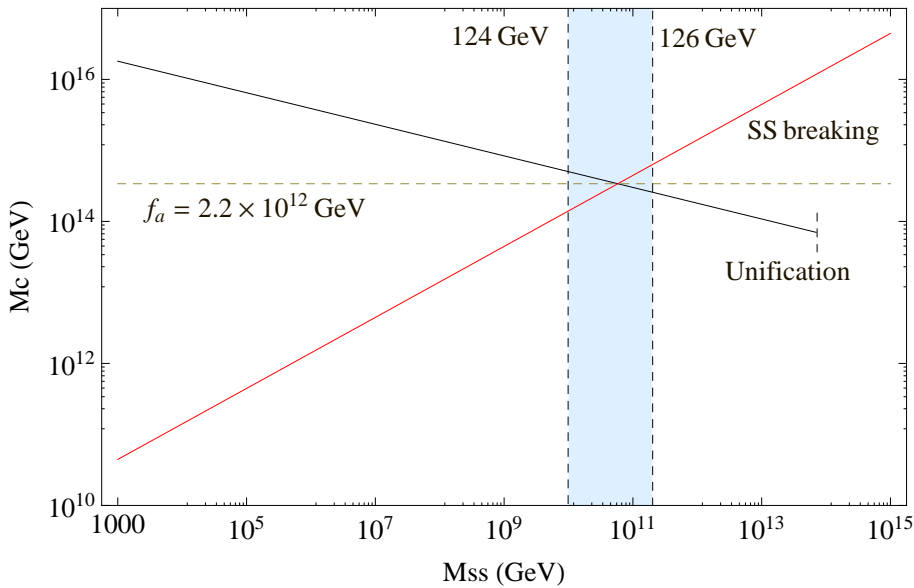}
\end{center}
\caption{Constraints on $M_{SS}$ and $M_c$ from gauge coupling unification
(black line) and  closed string flux induced SUSY breaking (red line).
The vertical (blue) band  shows the region of $M_{SS}$  at which the SM Higgs coupling $\lambda$ 
vanishes  for a Higgs mass in the range $124-126$ GeV and $m_t=173.2$ GeV, as 
extracted from ref.\cite{Giudice:2011cg,EliasMiro:2011aa}.
The horizontal line shows the value of the axion decay constant for the selected unification mass $M_c$.}
\label{cojoplot}
\end{figure}
%
This is displayed by the black line in figure \ref{cojoplot}. We have used the central values 
of the couplings
\beqa
\alpha_3(M_{EW}) \ & = & \ 0.1184 \pm 0.0007 \\
\alpha_{em}^{-1}(M_{EW})  \  & = & \ 128.91 \pm 0.02 \\
sin^2\theta_W(M_{EW}) \ & = &  \  0.23120 \pm 0.00015 \ .
\eeqa
One observes that 
one can get consistent unification for values of $M_{SS}$ up to slightly below 
$10^{14}$ GeV, which is required by the condition $M_{SS}<M_c$. The unification scale has also a lower bound at the same scale.
If however we impose that SUSY breaking is induced by closed string fluxes
with $M_{SS}= ( (2g_s)^{1/2}/\alpha_G^{1/2})\frac {M_c^2}{M_p}$ as explained in the
previous section, the values of both $M_{SS}$ and $M_c$ are fixed yielding
\beq
M_{SS} \ =\ 5\times 10^{10} \ GeV \ ;\  M_c\ =\ 3\times 10^{14} \ GeV \ .
\label{lasescalas}
\eeq
Thus one gets correct coupling unification consistent with closed string flux SUSY breaking
for SUSY broken at intermediate scale $\simeq 10^{11}$ GeV and a slightly low
$SU(5)$ unification scale of order $10^{14}$ GeV.
This immediately poses an apparent problem with proton decay that we will deal with in
section  6.

\begin{figure}[t]
\begin{center}
\includegraphics[width=6cm]{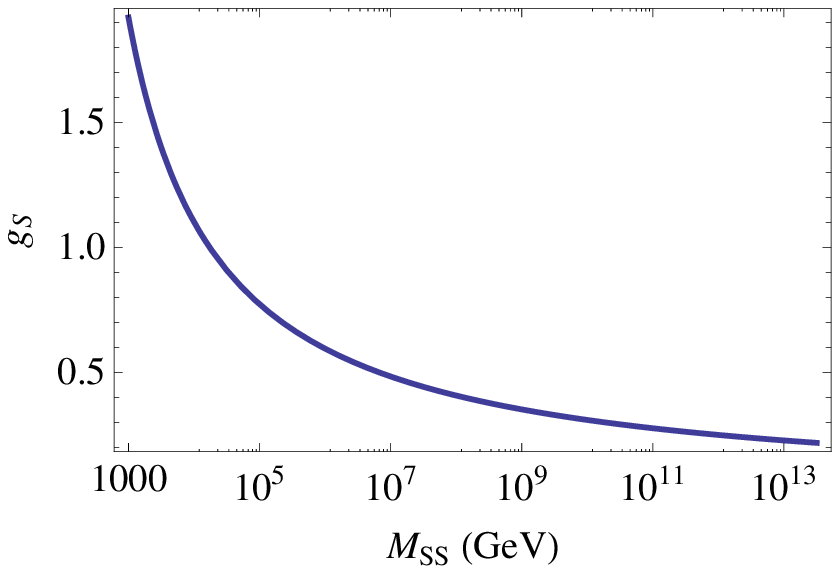} \ \ \
\includegraphics[width=6cm]{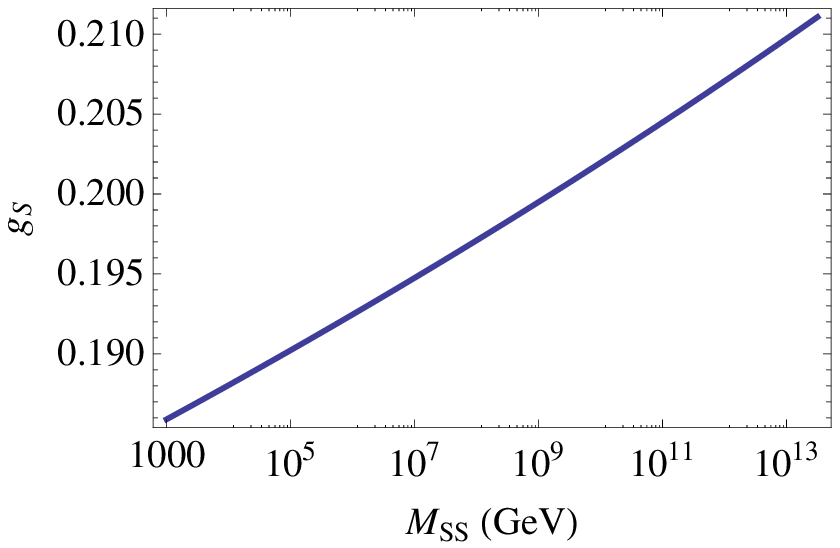} 
\end{center}
\caption{The string dilaton coupling constant versus $M_{SS}$ for consistent gauge
coupling unification. Left: With an MSSM content in the region
$M_{SS}-M_c$; Right:  With an additional vector-like triplet set $D+{\overline D}$ in that region.}
\label{gsgs}
\end{figure}

It is also interesting to display the value of $g_s$ as a function of $M_{SS}$ from eq.(\ref{calculogs}).
This is shown in fig. \ref{gsgs}. For the values in eq.(\ref{lasescalas}) one finds $g_s=0.28$. This shows that
the string coupling here is in a perturbative regime. On the other hand for 
values $M_{SS}\simeq 1$ TeV, corresponding to standard MSSM low-energy supersymmetry
one needs $g_s\gg 1$.
Note that in the context of F-theory the dilaton value $g_s$ varies over different locations in
extra dimensions and may be large or small, so both situations are possible in F-theory GUTs.

Another interesting point is that the unification line in fig. \ref{cojoplot} does not change if in the region 
$M_{SS}-M_c$ there are  {\it incomplete}  $SU(5)$ 5-plets, as equation (\ref{lineaunif}) does not change.
Thus for example the curve remains the same if the
$SU(5)$ partner of the Higgs doublets,  the triplets  $D,{\overline D}$ transforming like $(3,1,1/3)+c.c.$, remain in the spectrum
below $M_c$. These triplets are potentially dangerous since their exchange give rise to dimension 6 
proton decay operators.  The rate is above experimental limits unless $M_{D}\geq 10^{11}$ 
GeV \cite{Nath:2006ut}, see section 6.
 That is why in   GUTs one needs to perform some form of
doublet-triplet splitting so that the Higgs fields remain light but the triplets are superheavy.
 In our case however these triplets will get a 
mass of order $M_{SS}\approx 10^{11}$ GeV anyhow so they may be tolerated below $M_c$ and no
doublet-triplet splitting is necessary.  The presence of these triplets  does however affect the size  of the threshold corrections
and $g_s$. In this case one gets typically smaller $g_s$ which slowly grows as $M_{SS}$ increases, see fig. \ref{gsgs}.
For $M_{SS}\simeq 10^{11}$ GeV one gets 
$g_s=0.20$.

%

\section{The quartic coupling and the Higgs mass}

We have seen in the previous section how this ISSB framework 
is consistent both with gauge coupling unification and flux-induced SUSY breaking. Interestingly enough
it has been recently found that if a  non-SUSY SM Higgs is around
125 GeV, the SM RGE of the quartic self-coupling seems to 
drive it to a vanishing value at around $10^{11}$ GeV or so
(see e.g. \cite{Arbey:2011ab,Giudice:2011cg}).

\begin{figure}[t]
\begin{center}
\includegraphics[width=11cm]{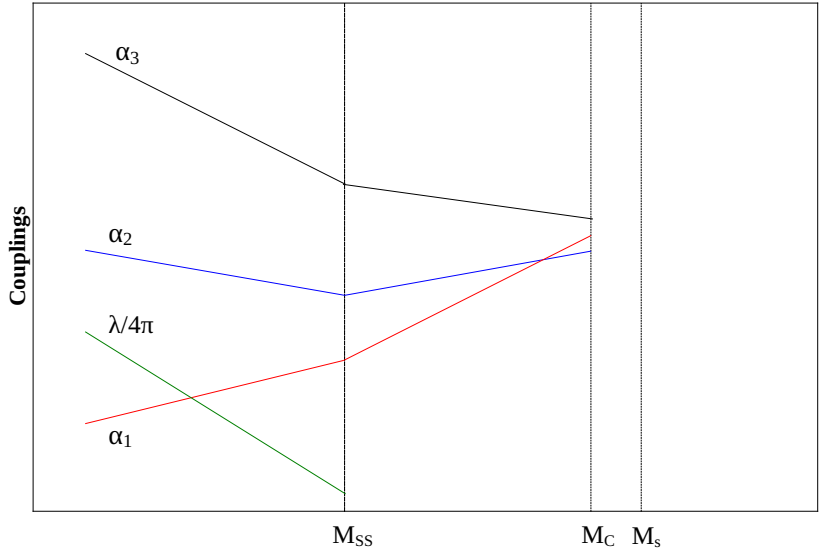}
\end{center}
\caption{Structure of scales and the running of the gauge and Higgs coupling constants in this
scheme.
}
\label{runningcouplings}
\end{figure}

The question is whether there is any SUSY/string based scheme in which that 
happens naturally. Note that the Intermediate Scale SUSY Breaking (ISSB)  described above corresponds to
a variant  of
the  High Scale SUSY Breaking (HSSB) scheme of Hall and Nomura in ref.\cite{Hall:2009nd}. 
This is just assuming a MSSM structure above a very large SUSY scale $M_{SS}$. All 
SUSY partners are heavy but there is still   some imprint left of the High Scale SUSY in
the Higgs sector. Indeed out of the two scalars $H_u,H_d$ in the MSSM only one linear
combination remains light below $M_{SS}$, i.e.
\beq
H_{SM} \ =\ sin\beta H_u \ -\  cos\beta H_d^*
\eeq
Then there is a quartic self-coupling $\lambda_{SS} |H_{SM}|^4$ with \cite{splitsusy,Hall:2009nd}
\beq
\lambda_{SS} \ =\ \frac {1}{8}
\left(g_2^2 \ +\ \frac {3}{5} g_1^2\right) \ cos^22\beta
\label{lambdasusy}
\eeq
which is inherited from the D-term scalar potential of the MSSM.
As we said  it has been shown
\cite{Giudice:2011cg,Arbey:2011ab,EliasMiro:2011aa,Holthausen:2011aa,Wetterich:2011aa,Degrassi:2012ry,Chetyrkin:2012rz,Bezrukov:2012sa}  that,  starting at low-energies with a SM Higgs with a mass around 124-126 GeV and running up the
 SM self-coupling $\lambda $ up in energies this coupling tends to zero 
 around a scale $10^{9}-10^{11}$ GeV (see fig. \ref{runningcouplings}).
 This would be consistent with the above High Scale SUSY Breaking scheme 
 if  at the scale $M_{SS}$ one had tan $\beta= \pm1$, so that $\lambda_{SS}(M_{SS})\simeq 0$.
 
 An interesting question is thus under what conditions one naturally gets tan $\beta\simeq \pm 1$.
 The general form of Higgs masses in the MSSM is\footnote{If in addition to the Higgs doublets 
 there remain Higgs triplets $D,{\overline D}$ below $M_c$, similar mass matrices will 
appear for them. However there will be no anthropic selection of light scalar triplets. 
So doublet-triplet splitting would be purely anthropic.}
 \beq
\left(
\begin{array}{cc}
 {{H_u}} \!\!\! & ,\  {{ H}_d^*}\\
\end{array}
\right)
\left(
\begin{array}{cc}
 { m_{H_u}^2} &   m_{3}^2\\   
  { m_3^2} & {m_{H_d}^2 }\\
\end{array}
\right)
\left(
\begin{array}{c}
  {{ H_u^*}} \\
  {{ H}_d}\\
\end{array}
\right) \,.
\label{matrizmasas}
\eeq
where we will take $m_3^2$ real. The condition for a massless eigenvalue is
$m_3^4=m_{H_u}^2m_{H_d}^2$. The massless eigenvector is then
\beq 
H_{SM} \ =\ sin\beta \ H_u \ \mp \ cos\beta \ H_d^*
\label{higgspm}
\eeq
with sin $\beta =\pm  |m_{H_d}|/\sqrt{m_{H_u}^2+m_{H_d}^2}$.
So in order to have a massless Higgs with tan $\beta\simeq \pm 1$
one needs to have the conditions
\beqa 
\label{fine-tuning1}
m_{H_u}^2  \	&  = & \  m_{H_d}^2  \\
m_3^4 \ &  =  &\  m_{H_u}^2m_{H_d}^2
\label{fine-tuning2}
\eeqa
We will take the negative sign in (\ref{higgspm}) from now on.
The first condition points to
an underlying symmetry under the exchange of $H_u$ and $H_d$, possibly slightly broken.
The second condition  does not necessarily imply any underlying symmetry, it is rather a fine-tuning constraint which has to
be there anyhow if we want to get a light Higgs. So this could be selected on anthropic grounds.

One can consider as a first option the direct construction of non-SUSY compactifications.
Examples of this could be e.g.  theories obtained from branes sitting at non-SUSY 
${\bf Z}_N$ orbifold singularities (see e.g., ref.\cite{Aldazabal:2000sa}). In this class of theories the particles in the
spectrum, typically involving both fermions and scalars, have no SUSY partners to start with.
These are however problematic since the spectrum generically 
contains tachyons from the closed string sector which destabilize the theory.
So we will restrict ourselves to theories in which there is an underlying $N=1$ SUSY
which is spontaneously broken to $N=0$. This will guarantee the absence of closed or open  string tachyons 
from the start.

There are several possible  sources for spontaneous SUSY breaking 
 in the IIB/F-theory  context  which may arise from open or closed string fluxes \footnote{In the F-theory context both 
closed and open string fluxes have a common origin as  $G_4$ fluxes.} which we discuss in turn.

  \vspace{0.5cm}

 {\it  i)SUSY breaking terms and  open string fluxes}

In Type IIB/F-theory in the large volume limit  the Higgs fields will
appear as KK zero modes.  
Open string fluxes, like the $F_a,F_Y$ mentioned above  are in general present
in order to generate chirality and symmetry breaking. These fluxes may induce also Higgs masses and 
SUSY-breaking terms as in eq.(\ref{matrizmasas}). We now review how such mass terms may appear 
in Type II toroidal orientifolds as discussed in \cite{Ibanez:2001nd}. In this reference a large class of
non-SUSY  Type IIA orientifolds with SM group and three chiral generations is discussed
in terms of D6-branes intersecting at angles. These models may be
converted into Type IIB orientifolds  with D7-branes by the duality that relates intersection angles $\theta_{ab}$
between two branes $a,b$ into magnetic fluxes at their overlap, through  the map
$\theta_{ab}= {\rm tan}^{-1} (2\pi\alpha' F_a)-{\rm tan}^{-1}(2\pi\alpha' F_b)$. 
In these non-SUSY models (see appendix A for a short review) the Higgs fields appear from the exchange of open strings
between an $SU(2)$ stack of branes $(b)$ and a brane $(c)$ or its orientifold mirror $(c^*)$.  The underlying torus is
factorized as ${\bf T}^2\times {\bf T}^2\times {\bf T}^2$ and the branes  $(b)$ and $(c),(c^*)$ are separated in the second torus
by a distance $Z_2^{(bc)}(Z_2^{(bc)^*})$. This means that the Higgs fields have a mass term proportional to this distance.
In addition the magnetic fluxes induce non-SUSY mass contributions. One gets a structure 
(in the dilute flux limit) \cite{Ibanez:2001nd}
\beqa
m_{H_u}^2\  & = & \ m_{H_d}^2 \ =\ \frac {Z_2^{(bc)}}{4\pi^2\alpha ' } \\
m_3^2 \ & = & \ |(F_b\ -\ F_c)^1-  (F_b\ -\ F_c)^3|
\eeqa
where $F_b^i,F_c^i$ are  fluxes from $U(1)$'s in the $b,c$  branes
going  through the first and third torus, see appendix A. Note that the first contributions correspond to
a $\mu$-term and that it automatically implies eq.(\ref{fine-tuning1}). This happens because one may 
understand $H_u$ and $H_d$  as coming from the  same $N=2$ hypermultiplet before SUSY is broken 
by the fluxes. On the other hand the size of $m_3^2$ depends on the size and alignment of fluxes. 
 For  $(F_b\ -\ F_c)^1= (F_b\ -\ F_c)^3$ one recovers unbroken  SUSY {\it  in the Higgs subsystem}.
 The $SU(2)\times U(1)_Y$ D-term quartic selfcoupling is given by
 $\lambda =(g_1^2+g_2^2)/8$  and, since this is a dimension 4 operator, it remains the same
 for $(F_b\ -\ F_c)^1\not= (F_b\ -\ F_c)^3$.
In principle for fixed $Z_2^{bc}$ one can fine-tune the 
fluxes so that eq.(\ref{fine-tuning2})  is met. Thus fine tuning of fluxes (or distance $Z_2^{(bc)}$) yields a massless
Higgs multiplet corresponding to tan$\beta = 1$.

Note that the open string flux misalignment corresponds to a D-term SUSY breaking. This means that by
itself this can only give SUSY breaking scalar masses but no gaugino masses (Higgssinos get SUSY masses for
$Z_2^{bc}\not=0$). Thus this class of SUSY breaking should be supplemented by further sources if
below $M_{SS}$ we want to get just the content of the SM. 

The above structure is generic and appears in any type II configuration where one can construct D-brane 
sectors with an $N=2$ hypermultiplet or a similar spectrum. In type IIB models with intersecting 
D7-branes or in F-theory GUTs such sectors arise quite naturally, since at the six-dimensional intersection 
of two 7-branes in flat space lives a 6d $N=1$ hypermultiplet that is equivalent to the 4d $N=2$ hypermultiplet 
of the construction above \cite{Berkooz:1996km}. Hence, in order to reproduce the above structure for the Higgs 
sector, one may consider the case where the Higgs matter curve $\Sigma_H$ yields a non-chiral, N=2 subsector 
of the theory.
As the presence of a net flux over a matter curve  induces a 4d chiral spectrum arising from it, the easier
way to preserve the N=2 structure is to impose that the integral of any relevant flux vanishes over $\Sigma_H$.
Note that in supersymmetric SU(5) F-theory models this option is usually not considered, since in order to achieve  
doublet-triplet splitting a net hypercharge flux is required to thread the Higgs curve(s). However, as mentioned
above in the present scheme we are not constrained by the amount of Higgs triplets at the scale $M_c$, and 
one may indeed consider the case where $\int_{\Sigma_H} F_Y =0$. 

In that case both $H_u$ and $H_d$ arise from the same curve $\Sigma_H$, and one may easily implement 
the mass structure (\ref{matrizmasas}). Just like for type IIA non-SUSY models, the term $m_3^2 H_u H_d^* + c.c.$
arises by inducing a non-vanishing D-term on this sector, which in this case is induced by worldvolume fluxes 
on $S$ which do not satisfy the condition $F \wedge J = 0$, $J$ being the K\"ahler form on $S$. For instance, 
let $\omega$ be the complex coordinate of $S$ along $\Sigma_H$ and $\omega_\perp$ the one transverse to it. 
Then if the flux felt by the doublets in $\Sigma_H$ is of the form $F = M_{||} d\omega \wedge d\bar{\omega} + 
M_\perp d\omega_\perp \wedge d\bar{\omega}_\perp $ the D-term condition reads $M_{||} + M_\perp = 0$ \cite{fi,afim}.
Hence, the off-diagonal terms in (\ref{matrizmasas}) read $m_3^2 = M_{||} + M_\perp$ and arise whenever such 
vanishing D-term condition is not met. 

Finally, the diagonal terms of the mass matrix (\ref{matrizmasas}) will correspond to a $\mu$-term. 
In a D7-brane setup dual to the toroidal models of appendix A this mass term appears by simply switching on
a continuous or discrete Wilson line along $\Sigma_H$. However, as mentioned before Wilson lines are typically 
not available in F-theory GUT models, and so the $\mu$-term cannot be generated by this mechanism. Instead, 
such supersymmetric mass term can be induced by the presence of closed string fluxes (see below) or at the 
non-perturbative level. Indeed, a $\mu$-term may appear at the non-perturbative level from string instanton 
effects \cite{stringinstantons} (see \cite{instantonrev,BOOK} for reviews).
Such $\mu$-terms are automatically symmetric under $H_u-H_d$ exchange and hence respect the
above structure. As we said, stringy instanton effects are of order exp$(2\pi/g_s)$ and  for $g_s\simeq 1/2$
could give rise to $\mu$-terms of the appropriate  order of magnitude $10^{-5} M_s$.

  \vspace{0.5cm}

{\it ii) SUSY breaking terms from  closed string fluxes and modulus dominance}

 Mass terms for scalar fields may also appear in the presence of closed string fluxes.
 Indeed this may be explicitly checked by plugging such backgrounds in the 
 DBI+CS action for 7-branes, see  \cite{softfromflux}. 
  In fact it is known that 
 SUSY breaking  {\it imaginary self-dual} (ISD)  IIB 3-form fluxes correspond  to 
 giving a non-zero vacuum expectation value to the auxiliary fields of Kahler moduli \cite{GKP}.
 So in order to see the effect of  closed string fluxes we will work here with the
 effective action and plug non-vanishing vevs for these auxiliary fields.
 
 In particular, in the context of Type IIB/F-theory compactifications a prominent role is
 played by the local Kahler modulus $T$ which is coupling to the  $SU(5)$ stack of 7-branes.
 In a general Type IIB/F-theory compactification this Kahler modulus is the one among 
 a number of such moduli which is relevant for the SUSY breaking soft terms, which will
 appear when $F_T\not= 0$. A good  model for this structure is considering the 
 CY manifold  ${\bf P}^4_{[1,1,1,6,9]}$ in ref.\cite{fernando} with one {\it small} Kahler modulus $T$ 
 and one {\it big} Kahler modulus $T_b$ with Kahler potential
\begin{equation}
K  \ =\ -2log(t_b^{3/2}\ -\ t^{3/2}) \ .
\label{kahlerswiss}
\end{equation}
with $t=2ReT$ and $t_b=2ReT_b$.
Here  one takes  $t_b\gg t$ and take both large so that the supergravity approximation is still valid. In the F-theory context
the analogue of these
moduli $t,t_b$ would correspond to the size of the 4-fold $S$ and the 6-fold $B_3$ respectively. 
For chiral matter fields living at F-theory matter curves one expects a behavior for the Kahler metrics
in the dilute flux limit \cite{cremades}
\begin{equation}
K  \ =\ \frac {t^{1/2}}{t_b} \ .
\label{metricswiss}
\end{equation}
On the other hand if the Higgs doublets  $H_u,H_d$ in  matter curves are not chiral,
they behave like scalars in a $N=2$ hypermultiplet, very much like in the previous case of
open string fluxes. Under these conditions one expects kinetic terms for the Higgs
multiplets of the form
\beq
\frac {t^{1/2}}{t_b}|H_u+H_d^*|^2 \ .
\eeq
This type of Higgs kinetic terms proportional to $|H_u+H_d^*|^2$  have been discussed in the past in the context
of heterotic orbifold compactifications with $N=2$ subsectors in the untwisted
spectrum  and they display a shift symmetry
under $H_u\rightarrow H_u+c, H_d\rightarrow H_d-c*$ \cite{Brignole:1996xb,Brignole:1995fb}.
Heterotic Type I S-duality indicates that such structure should  also be present in Type IIB 
orientifolds.
 Recently Hebecker, Knochel and Weigand  \cite{Hebecker:2012qp}
 have proposed that this shift symmetry may be at the origin of the 
 tan$\beta=1$ boundary condition and studied its appearance 
 also in Type II vacua.  In our context the assumption of T-modulus dominance 
 SUSY breaking allows to explicitly compute the relevant soft terms.
 Indeed applying standard supergavity formulae \cite{Brignole:1997dp}
 one obtains for  the Higgs mass parameters
 \beq
 m_{H_u}^2=m_{H_d}^2 = \frac {M^2}{2} \ ;\ \mu = -\frac {M}{2} \ ;\ m_3^2=\frac {3}{4} M^2
 \label{boundaries}
 \eeq
 so that 
 \beq
 m_{H_u}^2\ +\ \mu^2 \ =\ m_{H_d}^2\ +\ \mu^2=\ m_3^2\ =\ \frac {3}{4}  M^2
 \label{boundarybrignole}
 \eeq
 where $M=F_T/t$ is the gaugino mass, with $F_T$ the auxiliary field in the $T$ chiral
 multiplet. Now, unlike the open string flux case, 
  the diagonal masses have both a SUSY contribution
 and a SUSY-breaking contribution and there is automatically a
 massless Higgs boson.  We  again obtain tan $\beta=1$ at the unification scale,
 this time automatically due to the mentioned shift symmetry. This value is however 
 renormalized, as we point out below.

%

As a general conclusion, we see that in string theory models in which the Higgs sector
corresponds to  a $N=2$ subsector with $H_u,H_d$ sitting in a hypermultiplet (before SUSY breaking),
the condition $m_{H_u}^2\simeq m_{H_d}^2$ is naturally obtained. In addition off-diagonal $m_3^2$ terms
may be induced both by effects from open and closed  string fluxes.

Let us finally comment on possible generalizations of this minimal Higgs structure.
A first generalization is starting with $n_H$ sets of Higgs particles above  $M_{SS}$.
In that case minimal landscape fine-tuning will still prefer that only 
one combination of the  $2n_H$  Higgs scalars  remains massless. Depending on 
how the original Higgs multiplets coupled to the different families the resulting Yukawas
could inherit an interesting  flavor structure. 
Another possible extension could be to dispose of R-parity in the initial MSSM spectrum
  since L/B-violating dimension four operators are suppressed due to the large mass of
  sfermions. In this case the Higgs $H_d$ could mix with sleptons $L_i$. However in
this case the approximate symmetry under the exchange of $H_u$ and $H_d$ 
would typically be  absent and the prediction tan $\beta\simeq 1$ would be in danger, so this
particular bilinear should be slightly supressed. 

\section{Higgs mass fine-tuning}

We have seen how one may naturally obtain a massless Higgs with tan$\beta=1$ in
string theory and, in particular, also  in  the
context of Type IIB  constructions with mass terms induced by open and closed string
fluxes. In general one has to fine-tune the parameter $m_3^2$ with the Higgs masses 
$m_{H_u}^2=m_{H_d}^2$. This may be done e.g. by partially  canceling the  contributions to $m_3^2$ 
from open and closed string fluxes.

However the above results are subject to loop corrections which will force to a redefinition
of the fine-tuning. If the SUSY breaking scale is of order $M_{SS}\simeq 10^{11}$ GeV the fine-tuning 
should be done to at least 4-loop order to  cope with  a hierarchy of nine orders of magnitude down to the EW scale.
Still the idea is that even after these further fine-tuning corrections the Higgs scalar which remains light 
is approximately the one corresponding to the combination $H_u- H_d^*$, i.e. that approximately
tan$\beta\simeq 1$. 

In fact, if we assume that tan$\beta (M_{SS})=1$ (as e.g.  in eqs.(\ref{boundarybrignole})) before
 any loop correction is included, we know that the 
running of the parameters in between the scales $M_c$ and $M_{SS}$ will renormalize tan$\beta$. We expect that the
large top quark Yukawa coupling will make $m_{H_u}^2<m_{H_d}^2$ after loop corrections. We also expect to obtain 
one massive Higgs eigenstate and a second one slightly tachyonic. This may be compensated to get a massless 
Higgs boson at this level by tuning with an open or closed string flux as explained in the previous section. 
Still, after this fine-tuning, the value of tan$\beta$ is no longer 1 but is given by tan$\beta =|m_{H_d}(M_{SS})|/|m_{H_u}(M_{SS})|$,
as explained in the previous section.   In addition to this there will be higher order finite   loop corrections which are expected to
yield smaller negligible contributions to tan$\beta$.  So a good estimation of tan$\beta$ at the scale $M_{SS}$ should
be given by taking into account the running of the parameters in between $M_c$ and $M_{SS}$.

To compute the  value of tan$\beta$ at $M_{SS}$ we have to consider the RGE for the MSSM parameters in the 
region $M_{SS}-M_c$. In the present case we know that with a single Higgs field at the electroweak scale 
only the top-quark Yukawa coupling $h_t$ is relevant in this equation. Fortunately, the one-loop RGE in the
$h_t\gg h_b,h_{\tau}$ limit were solved analytically in ref.\cite{ilm} for the case of universal soft terms, i.e.
as in the CMSSM model, which should be more than enough to evaluate this renormalization effect.
One has tan$\beta(M_{SS})=|m_{H_d}(M_{SS})|/|m_{H_u}(M_{SS})|$ with
\beq
m_{H_d}^2(t) \ =\ m^2 \ +\ \mu^2 q^2(t)\ +\ M^2 g(t)
\label{ilm1}
\eeq
\beq
m_{H_u}^2(t)\ =\ m^2(h(t)-k(t)A^2)\ +\ \mu^2q^2(t)\ +\ 
M^2e(t)\ +\ AmMf(t)
\label{ilm2}
\eeq
where $m,M,A,\mu $ are the standard universal CMSSM  parameters  at the unification scale $M_c$, $t=2log(M_c/M_{SS})$  
and $q,g,h,k,e,f$ are known functions of the top Yukawa coupling $h_t$ and the three 
SM gauge coupling constants. For completeness these functions are provided in Appendix B.
Note that in order to compute tan$\beta (M_{SS})$ we do not need to know how $m_3^2$ runs since its value is
fixed by the fine-tuning condition $m_3^4=m_{H_u}^2m_{H_d}^2$ at $M_{SS}$.

\begin{figure}[t]
\begin{center}
\includegraphics[width=11cm]{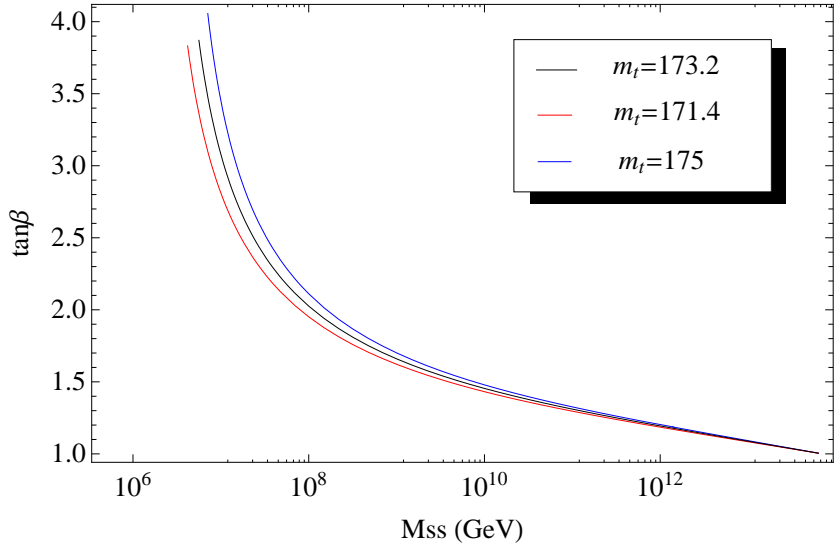}
\end{center}
\label{plotillouno}
\caption{Renormalization of tan$\beta$ in the region $M_{SS}-M_c$ as a function of $M_{SS}$
for three values of the top quark mass.}
\end{figure}
%

These functions involve integrals of coupling constants  over the region $M_{SS}$ to $M_c$. 
There is an explicit dependence on the universal soft terms $m,M,A,B,\mu$ which are all of order
$M_{SS}$ but the results are quite insensitive to the precise values of those parameters. 
For definiteness we have computed  tan$\beta (M_{SS})$ for the boundary conditions
$m^2=M^2/2$, $A=-3M/2=B$, $\mu=-M/2$, with $M=M_{SS}$, corresponding to the modulus dominance
SUSY-breaking soft terms described around eq.(\ref{boundaries}), see e.g.  \cite{aci1}.
In figure \ref{plotillouno} 
 we show the
dependence of tan$\beta(M_{SS})$ as a function of $M_{SS}$, where $Mc$ is taken as  in section 3,
from the gauge coupling unification condition. One observes that for
$M_{SS}$ in the range $10^{8}-10^{12}$ GeV tan$\beta$ is only slightly increased to a 
value around tan$\beta \simeq 1.2-1.4$,  depending on the value of the top-quark mass.
In fact using the above formulae one can expand in a power series of the square of the
top Yukawa coupling to find
\beq
{\rm tan}\beta(M_{SS}) \ =\ 1.00 \ +\ h_t^2(M_c)\times 0.58 \ + \ .... \ .
\eeq
It seems then that the tree level value of tan$\beta$ is only slightly deformed away from 1 after
loop corrections. As we said, higher loop effects required to do a fully consistent fine-tuning 
are not expected to spoil this conclusion. An analogous conclusion was reached in \cite{Hebecker:2012qp}
using different methods.\footnote{These results remain unchanged in the case of a R-parity 
violating MSSM since the new B/L-violating couplings will only appear in the Higgs mass running at
two loops.} One can trivially extend  the calculation to the case in which color triplets 
$D,{\overline D}$ remain below $M_c$ with very similar results.

A natural question is whether the  Higgs mass terms discussed above {\it scan} in the string landscape. 
These masses depend on the local value of  string {\it flux densities} in the region  in the
compact dimensions where the SM fields are localized. These  local densities are in general not quantized,
it is their integrals over  3- and 2-cycles which are  quantized. As is well known in a generic compactification 
there may be of order a hundred different  quantized closed string  fluxes which may be turned on. All of them in general 
may contribute to the cosmological constant and could play a role in its  anthropic solution \cite{weinberg}. The required energy
spacing for the c.c. constant is so minute that at least some of these fluxes should e.g. be combined with anti-D3-branes 
on CY-throats in order to be able to fine-tune the c.c. following the KKLT approach
\cite{kklt}. On the other hand only 
a selected number of fluxes affect the SM branes in the cycle $S$. Again although these fluxes are quantized
it is the density $G_3$ at the location of the 7-branes which is relevant. However varying the flux quanta one can also
control this local density. So, indeed, it seems plausible that the subset of the fluxes  going through  $S$
will scan in the string landscape. It would be interesting to materialize in some detail this expectation.

\section{Proton decay}\label{proton}

As we already advanced with a unification scale as low as $M_c=3\times 10^{14}$ GeV 
there is a danger of dimension 6 operators giving rise to proton decay rates much faster than
experiment. In standard field theory GUTs,
the proton decay dim=6 operators obtained after integrating out the massive $X,Y$  doublet of gauge bosons are
\cite{Nath:2006ut}
\beqa
O_1 \ & =& \  \frac {4\pi \alpha_G}{2M_{X,Y}^2} {\overline {U_{aL}^c}} \gamma^\mu Q_{aL}  {\overline {E_{bL}^c}}\gamma_\mu Q_{bL} \\
O_2 \ & =& \  \frac {4\pi \alpha_G}{2M_{X,Y}^2} {\overline {U_{aL}^c}} \gamma^\mu Q_{aL}  {\overline {D_{bL}^c}}\gamma_\mu L_{bL}  \ .
\eeqa
The first operator  arises from the exchange of the heavy gauge bosons with masses 
$M_{X,Y}$  between two 10-plets whereas the second
from the exchange between a 10-plet and a 5-plet.  
Experimentally, 
the Super-Kamiokande limit on the  chanel $p\rightarrow \pi^0 e^+$ gives an absolute lower limit
$\tau_p>5\times 10^{33}$ years \cite{Nishino:2009aa}. This corresponds to a bound on $M_{X,Y}$ 
\beq
M_{X,Y}\ \geq \sqrt{\frac {\alpha_G}{1/39}} \ 1.6\times 10^{15} \ GeV
\eeq
A value $M_{X,Y}=M_c=3\times 10^{14}$ GeV is 5 times smaller and that could pose a problem.
In F-theory GUTs the same proton decay operators as above will appear, the difference now being that 
the symmetry is broken due to a hypercharge flux. Due to this fact the coefficients of the operators may
change substantially, as we now discuss.

Indeed, considering proton decay in the context of F-theory SU(5) unification provides a new 
interesting mechanism to suppress proton decay. A microscopic computation of the above dimension 6 proton decay operator would involve first computing couplings of the form e.g. ${\overline {U_{aL}^c}} X_\mu Q_{aL}$ and then integrating out the massive doublet $X,Y$. The computation of such trilinear couplings is rather similar to the computation of Yukawa couplings, in the sense that it also involves a triple overlap of internal wavefunctions, namely
 \beq
 \Gamma_1^{ij} \ =\ 2\, m_* \int_S (\Psi_{10}^i)^\dag \Psi_{10}^j \Phi_{X,Y} \quad \quad \Gamma_2^{ij} \ =\  2\, m_* \int_S (\Psi_{\bar 5}^i)^\dag \Psi_{\bar 5}^j\Phi_{X,Y} 
\label{d3coups}
\eeq
where now $\Phi_{X,Y}$ are the internal wavefunctions of the broken SU(5) bosons $X,Y$. These form a doublet of
massive gauge bosons with quantum numbers $(3,2,5/6)+c.c.$.

In standard 4d GUTs, the value of such couplings does not depend on the vev of the Higgs in the ${\bf 24}$ of SU(5), and so it is exactly the same before and after SU(5) breaking (to leading order).
 Hence, one may extract the trilinear couplings like ${\overline {U_{aL}^c}} X_\mu Q_{aL}$ directly from the SU(5) Lagrangian as the strength by which SU(5) gauge bosons couple to chiral matter, namely $(4\pi\alpha_G)^{1/2}$. 

Now, the key point for proton decay suppression in F-theory is the fact that the ingredient that triggers SU(5) breaking is not a vev for a scalar in the adjoint of SU(5), but the presence of the hypercharge flux $F_Y$ along the GUT 4-cycle $S$. The mass of the 
$X,Y$ gauge bosons is given by
\beq
M_{X,Y}^2 \ =\ \frac {5\mu }{6\pi}
\eeq
where $\mu=\sqrt{N_Y^2+\tilde N_Y^2}$ measures the density of hypercharge flux (see appendix \ref{ap:proton}), which we take constant for simplicity. The flux quantization condition implies that  $5/3(F_Y/2\pi)$ is quantized in $S$  (i.e., its integral over 2-cycles of $S$ is an integer), 
so that $N_Y, \tilde{N}_Y\approx 6\pi/5{\rm Vol}_S^{-1/2}$ and indeed $M_{X,Y}\simeq M_c\simeq {\rm Vol}^{-1/4}$.
Finding the wavefunctions in (\ref{d3coups}) involves solving a Dirac or Laplace equation for them, in which any flux threading $S$ will enter. We then have that both the wavefunctions for chiral fields and massive gauge bosons $X$, $Y$ depend on the internal fluxes on $S$, and in particular on the hypercharge flux $F_Y$. As a result, adding an hypercharge flux will necessarily change the value of the effective 4d couplings (\ref{d3coups}): while in the absence of $F_Y$ such couplings must be $\propto \alpha_G^{1/2}$ in its presence they will have a new value.

To show that this new value will be suppressed with respect to $\alpha_G^{1/2}$ we need some machinery from wavefunction computation in F-theory GUT models. Here we will try to be schematic, referring the reader to appendix C and to \cite{eterno} 
(see also \cite{fi,afim,yukothers,corryuk,cdp}) for more details on the subject. In F-theory SU(5) models there are basically two kinds of wavefunctions: the ones that are peaked at the matter curves of $S$, namely $\Psi_{10}^i$, $ \Psi_{\bar 5}^j$ and $\Phi_{H_{U,D}}$, and the ones that are spread all over the 4-cycle $S$, namely the SU(5) gauge bosons and in particular $\Phi_{X,Y}$.  As they come from different sectors of the theory, these two kinds of wavefunctions feel the effect of the hypercharge flux in a different way. 

Indeed, let us consider the wavefunctions involved in the coupling $\Gamma_{1}$ in (\ref{d3coups}). Solving for them in a local patch of $S$ and assuming that the 4-cycle $S$ is sufficiently large  (see appendix C and \cite{eterno} for more details) we have that 
\begin{eqnarray}
\Psi_{10}^{i}& =& 
\left(
\begin{array}{c}
0 \\ \vec{v}
\end{array}
\right)\,
\psi^i_{10},
\quad 
\psi^i_{10} \, =\,\gamma_{10}^{i} \, m_*^{4-i} \, x^{3-i} \ e^{-\frac{|M_{x}+q_Y\tilde{N}_Y|}{2}|x|^2} e^{- m^2 |y|^2 - q_S {\rm Re} (x\bar y)}
\label{wf10}\\
\Phi_{X,Y} & = &\gamma_{X,Y}\, m_*\, e^{-\frac{5}{12} \mu (|x|^2+|y|^2)  }
\label{wfXY}
\end{eqnarray}
where $(x,y)$ stand for local complex coordinates of the 4-cycle $S$, and we have assumed that matter curve supporting the chiral fields ${\bf 10}$ is given by $\Sigma_{10} = \{ y=0\}$. The hypercharge dependence of the wavefunction $\Psi_{10}$ is encoded in the hypercharge value $q_Y$ and in $q_S = N_F + q_Y N_Y$, so that for a non-vanishing $F_Y$ particles with different hypercharge have different wavefunctions. Here $M_x$, $\tilde{N}_Y$, $N_Y$ and $N_F$ stand for densities of fluxes threading the 4-cycle $S$, and in particular $M_x$ is the density of the flux necessary to have three families of {\bf 10}'s along $\Sigma_{10}$.  The parameter $m^2$ stands for the slope of the intersection between the SU(5) 4-cycle $S$ and the U(1) 7-brane intersecting $S$ in $\Sigma_{10}$. Such intersection scale is typically of the order of the fundamental scale of F-theory $m_*$ ($\simeq M_s$ in a perturbative IIB orientifold),
which implies that $\Psi_{10}^i$ are highly peaked along the matter curve $\Sigma_{10} = \{ y=0\}$.  Finally, $\vec{v}$ is a three-dimensional vector that depends on $m^2$ and the flux densities, and the $\gamma$'s are normalization factors that insure that such fields are canonically normalized. 

Both $\vec{v}$ and the quantities that appear in the exponential factor of $\psi_{10}^i$ are family independent: the only dependence of the family index $i$ corresponding to the power of $x$ (the matter curve $\Sigma_{10}$ coordinate) that appears in the wavefunction. It has been found  \cite{fi,afim,yukothers,corryuk} that with this prescription (that assigns the power $x^2$ to the first family, etc.) one can reproduce the mass hierarchy between families observed in nature. 

Notice that the fact that $M_x$, $\tilde{N}_Y$ and $m^2$ are non-zero gives a gaussian profile to these wavefuctions, and this allows to carry the integral for $\Gamma_1$ by replacing $S$ with $\IR^4$. This is important since otherwise we would need geometrical information 
about the full manifold $B_3$, which is in general not available.
 Notice also that the wavefunction for the boson $X, Y$ is only affected by the hypercharge flux density $\mu$, and that in the limit $\mu \raw 0$ we recover a constant wavefunction. This is to be expected, since at this limit the SU(5) symmetry is restored and $X,Y$ become massless gauge bosons, which always have a constant profile. 

Given these facts we are now ready to compute the coupling $\Gamma_1$ above. First notice that in the limit $\mu \raw 0$ the integral is trivial in the sense that $\Phi_{X,Y} = \gamma_{X,Y} m_*$ is constant, since
\beq
2 m_* \int_S (\Psi_{10}^{i})^\dag \Psi_{10}^{j} \Phi_{X,Y}  = 2 \gamma_{X,Y} m_*^2 \int_S (\Psi_{10}^{i})^\dag \Psi_{10}^{j} \approx \alpha_G^{1/2} \delta^{ij} 
\label{initial}
\eeq
where  used that for $\mu = 0$, the normalization factor is simply $\g_{X,Y} = {\rm Vol}_S^{-1/2} m_*^{-2}\approx   \alpha_G^{1/2}$. Hence in this limit we recover the result expected from SU(5) gauge invariance. 

This result is no longer true when $\mu \neq 0$ and so the wavefunction $\Phi_{X,Y}$ has a non-trivial profile. Then one finds that there is a suppression in the above coupling which is family dependent, and bigger for lower families. Indeed, to get an estimate of this coupling it is useful to take the approximation $m^2 \sim m_*^2 \gg M_x, \mu$ and treat the Gaussian profile exp$(-m^2 |y|^2)$ as a $\delta$-function in the coordinate $y$, which is nothing but asking that the matter wavefunctions $\Psi_{10}^i$ are fully localized in $\Sigma_{10}$. That is, we take the limit $m^2 \raw \infty$ in which 
\beq
(\psi_{10}^i)^* \psi_{10}^j \, \raw \, \gamma_{10}^i \gamma_{10}^j m_*^{8-i-j} \bar{x}^{3-i}x^{3-j}  \, e^{-|M_x+ \bar{q}_Y \tilde{N}_Y| |x|^2} \frac{\pi}{m^2} 
\delta(y)
\label{product}
\eeq
and so the integral must be basically taken over $\Sigma_{10}$. Here $\bar{q}_Y = (q_{Y_{p}} + q_{Y_{q}})/2$ is the mean value of hypercharge for the two particles of the 10-plot participating in the amplitude. Taking into account that in this limit the normalization factors are \cite{eterno}
\beq
\gamma_{10}^i \,=\, \frac{1}{\sqrt{2 (3-i)}\pi} \left(\frac{|M_x + q_Y \tilde{N}_Y|}{m_*^2}\right)^{\frac{4-i}{2}}\quad \quad 
\gamma_{X,Y}\, =\, \frac{1}{\sqrt{2}\pi} \frac{5\mu}{6 m_*^2}
\eeq
we obtain that
\beqa\nonumber
2 m_* \int_S (\Psi_{10}^{i})^\dag \Psi_{10}^{j} \Phi_{X,Y} &  = & {\delta^{ij}} \frac{5\mu}{6\sqrt{2}\pi m_*^2} \ \left( \frac{|M_x + q_{Y_p} \tilde{N}_Y|^{1/2}|M_x + q_{Y_q} \tilde{N}_Y|^{1/2}}{|M_x + \bar{q}_Y \tilde{N}_Y| +\frac{5}{12} \mu}\right)^{4-i} \\ 
& \approx & \delta^{ij} \alpha_G^{1/2} \left( \frac{|\sigma^2 + (\frac{5}{12} \tilde{N}_Y)^2 |^{1/2}}{\sigma+\frac{5}{12} \mu}\right)^{4-i}
\label{final}
\eeqa
where we have defined $\sigma = |M_x + \bar{q}_Y \tilde{N}_Y|$ and used $|q_{Y_p} - q_{Y_q}| = 5/6$ and  $\mu \approx 6\pi /5 {\rm Vol}_S^{-1/2}$. This result is reproduced in appendix C without taking the $\delta$-function approximation.

Since $\mu > \tilde{N}_Y$, the coupling (\ref{final}) is indeed suppressed with respect to the 4d GUT result $\alpha_G^{1/2}$, and that the suppression is bigger the lighter the family. Since we are interested in proton decay operators one could in principle  focus 
on the first family $i=1$, in which by assuming  $M_x \approx N_Y5/12 \approx \tilde{N}_Y5/12$ we already obtain a suppression factor of  around 1/5, and much bigger if $N_Y > \tilde{N}_Y$. In fact, being more rigorous, we would really need to take into account
the fact that the actual physical first generation wave functions  will be proportional to a linear combination of the $x^2,x,1$ monomials.
Even if this extra terms are present, 
one expects  the first generation to be dominated by the $x^2$ monomial with a small contamination (related to
mixing angles) from the other two.\footnote{We thank P. C\'amara for discussions on these points.}  In any event, the presence 
of a suppression will be generic.

The fact that the suppression factor is bigger for each family can be given an intuitive understanding, since in F-theory families with smaller Yukawa couplings are those that have a higher polynomial degree $x^n$ in their wavefunction (see eq.(\ref{wf10})). Such higher power gives a compensating effect to the localization that arises from the family independent exponential factor exp$(-a|x|^2)$, that tends to localize the triple overlap around $x=0$. The lighter the family the bigger the compensating effect, thus the smaller the coupling. 

This understanding of the coupling strength in terms of exponential factors gives yet another mechanism for suppressing the dimension six proton decay operators. Indeed, notice that in (\ref{wfXY}) we have described the wavefunction for the massive $X,Y$ bosons in terms of a Gaussian function on $S$ peaked at $x=y=0$. However, that the wavefunction $\Phi_{X,Y}$ peaks there is in fact a choice that we have made biased by the local patch description of our F-theory model setup. Unlike for the wavefunctions $\Psi_{10}^j$, whose equations of motion force them to be localized at the matter curve $\Sigma_{10} = \{y=0\}$, there is nothing special about $y=0$ for the wavefunctions of the gauge bosons $X,Y$ which only depends on the hypercharge flux $F_Y$ and on the geometry of the 4-cycle $S$. Only these two factors will determine where the peak of the wavefunction $\Phi_{X,Y}$ is, so there is a priory no reason  to think that it will be peaked at any matter curve. Now, if the wavefunction $\Phi_{X,Y}$ is not peaked at $y=0$ but somewhere else the $\delta$-function in (\ref{product}) will yield an extra suppression upon integration on the complex coordinate $y$, as the wavefunction density for $\Phi_{X,Y}$ will be exponentially suppressed away from its peak.

\begin{figure}[t]
\begin{center}
\includegraphics[width=11cm]{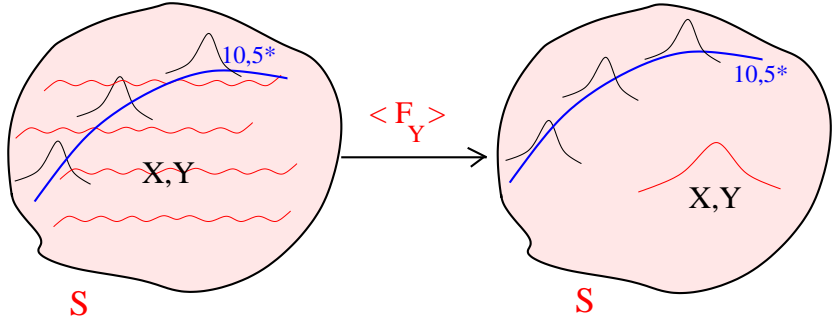}
\end{center}
\caption{Coupling of  $SU(5)$ off-diagonal gauge bosons 
$X,Y$. Before symmetry breaking by hypercharge fluxes the wave
function of $X,Y$ is extended over the whole 4-cycle $S$. After the hypercharge 
flux $F_Y$ in introduced their wavefunction is localized and their coupling 
to $10,{\bar 5}$ fields is supressed.
}
\label{dianadegales}
\end{figure}
%

To summarize, F-theory SU(5) models have naturally suppressed dimension 6 proton decay operators, because the mechanism that breaks the SU(5) symmetry - the hypercharge flux $F_Y$ - also affects the couplings where these operators come from. Indeed, the presence of the hypercharge flux deforms the wavefunction profile for the fields ${\bf 10}$, ${\bf \bar{5}}$ and $X,Y$, as illustrated in figure \ref{dianadegales}. In particular it affects the $X,Y$ bosons, which instead of being massless gauge bosons extended evenly over the whole 4-cycle $S$, are due to $F_Y$ massive modes peaked at some point of it. Such localization effect indeed changes the value of the couplings (\ref{d3coups}) as we have shown above in a simplified computation reproduced in more detail in Appendix C. Moreover, for the sake of simplicity we assumed above that the peak of the $X,Y$ wavefunction lied on top of the matter curve $\Sigma_{10}$ where the 10-plet resides. There is no reason for this assumption to hold in a global description of our setup, so the $X,Y$ wavefunction will in general be suppressed in the region of $\Sigma_{10}$ and there will be a further suppression to the coupling of $X,Y$ to quarks-leptons. It is easy to see that  any of these suppression mechanisms allow to have a rate for proton decay consistent with experimental limits. 
Note  however that the precise value of the coefficient of the operators depends
on the details (i.e. local fluxes) of the model. Still these results allow for the possible detection of proton decay 
through e.g.  the channel $p\rightarrow \pi^0 e^+$, typical of non-SUSY unification, in future proton decay experiments.

If  Higgs triplets  $D,{\overline D}$ with a mass $M_D\simeq M_{SS}\simeq 10^{11}$ GeV
are present in the spectrum, there will appear additional contributions to proton decay close to the present experimental limits
\cite{Nath:2006ut}.
They would come from the exchange of the scalar fields ${\tilde D},{\tilde {\overline D}}$ among quarks 
and leptons of the first and second generations from Yukawa couplings, with $p\rightarrow \mu^+K^0$ 
the dominant channel.  In field theory GUTs these Yukawa
couplings are directly related to the Yukawa couplings of the Higgs doublets due to the $SU(5)$ symmetry. 
In our case however the relevant D-field Yukawas are different to those of the Higgs, again due to the presence of
the hypercharge flux \cite{eterno}. One still expects those Yukawas to be of the same order of magnitude, i.e. of order 
$10^{-5}$ for the first generation.  The combination of a massive D-field with the smallness of  Yukawa couplings 
make  these extra dimension 6 contributions compatible with experimental bounds, given the uncertainties.

Note in closing  that dimension 5 proton decay operators are very much suppressed in the present framework 
due to the large mass of the SUSY partners. Additional sources of proton decay could appear if the underlying 
MSSM contains dimension 4 R-parity violating couplings. These could give rise to new dimension 6 operators 
by the exchange of sfermions but the rate will be again suppressed by the large mass of the SUSY partners combined with
the expected smallness of the R-parity violating couplings involving the first generations.

\section{Other consequences}

\subsection{Axions}

The strong CP problem is a naturality problem with no obvious anthropic solution. In this sense it is quite 
satisfactory that string theory has natural candidates for the axion solution of the strong CP problem.
As shown in eq.(3.1)  the imaginary part of the local K\"ahler modulus Im $T$ has axionic 
couplings to the QCD gauge bosons, and hence is in principle an axion candidate which could solve the
strong CP problem \footnote{The $\tau$ complex dilaton scalar has also axionic couplings
but Im$\tau$ gets generically massive in the presence of closed string fluxes.}.   In the Type IIB/F-theory scheme under discussion 
it is an important point the decoupling of the local  GUT physics sitting on the local $S$ 
4-cycle from the  global physics of the full six extra dimensions. In particular it is the 
local Kahler modulus $T$ which couples to the  $SU(5)$ gauge bosons as shown in 
eq.(3.1).  
One can compute the associated axion scale $F_a$ from the kinetic term of
the modulus $T$ (see e.g.  \cite{BOOK})
\beq
F_a^2\ =\ \frac {M_p^2}{4\pi (8\pi^2)^2}\frac {\partial K(T,T^*)}{\partial T\partial T^*} \ =\ 
\frac {M_p^2}{4\pi (8\pi^2)^2}\frac {3t^{-1/2}}{8t_b^{3/2}}
\eeq
where in the last equality we have used eq.(\ref{kahlerswiss}), which correctly
features the decoupling of the local $SU(5)$ physics from the global properties
of the compact manifold.
For the local modulus  one has $t=1/\alpha_G$ and 
\beq
t_b\ =\ \frac {V_6^{2/3}}{g_s\alpha'^2(2\pi)^4} \ =\ 
\frac {\alpha'^{1/2}g_s^{1/4}}{\sqrt{8}}M_p
\eeq
where in the last equality we have used eq.(\ref{masaplanck}). Using eq.(\ref{compactmas})  one finally
obtains
\beq
F_a \ =\ \left(\frac {18}{\pi ^2}\right)^{1/4}\ \frac {M_c}{16\pi^2} \ .
\eeq
Note that the axion decay constant is directly related to the compactification scale 
(or the string scale via eq.(\ref{compactmas})) and hence may be naturally low. 
This is to be contrasted to the heterotic model-independent axion Im$S$ whose
axionic coupling is directly tied to the Planck scale through 
$F_a^{het}=\alpha_GM_p/(8\pi ^{3/2})\simeq 10^{16}$ GeV (see e.g. \cite{BOOK} and references
therein).
In our case, for the preferred value $M_c=3\times 10^{14}$ GeV one obtains
\beq
F_a \ \simeq \ 2\times 10^{12} \ GeV \ .
\label{faultimo}
\eeq
This is an interesting value since $F_a$ it is in the allowed 
QCD invisible axion range. It is at the upper limit of the allowed
window, which is in fact required for the axion to be a viable
dark matter candidate.  This is also fortunate because in this scheme there are
no light neutralinos as in the MSSM or split SUSY which could play the role of
dark matter.

The mass of the axion is given through  standard formulae  by  (see. e.g.\cite{PDG})
\beq
m_a\ =\ \frac {z^{1/2}}{(1+z)}\frac {f_\pi m_\pi }{F_a}\ =\ \frac {0.6\times 10^3\ \mu eV}{F_a/(10^{10}GeV)}
\eeq
where we have taken $z=m_u/m_d=0.56$. For the $F_a$ value in (\ref{faultimo}) one gets an
axion mass
\beq
m_a\ \simeq \  2.7 \ \mu eV \ .
\eeq
Due to the underlying $SU(5)$ symmetry the coupling of the axion to
photons is directly related to $F_a$ by a factor $sin^2\theta_W=3/8$ (this is analogous to the DFSZ 
axion case \cite{DFSZ}).  In particular, defining the 
(normalized) axion-photon coupling as
\beq
\frac{G_{a\gamma \gamma}}{4}\ a\ F^\gamma\wedge F^\gamma
\eeq
one obtains 
\beq
G_{a\gamma \gamma}\ =\  \frac {\alpha_{em}}{2\pi F_a}(\frac {8}{3}\ -\ \frac {2}{3} \frac {(4+z)}{(1+z)})\ \simeq 0.38  \times 10^{-15}(GeV)^{-1} \ .
\eeq
These values are not far from the limits obtained from 
searches with the microwave cavity experiment
ADMX 
for  cosmic axion dark matter \cite{ADMX}. They obtain
\beq
\frac {|G_{a\gamma \gamma}|}{m_a/\mu eV} \ <\  5.7\times 10^{-16} (GeV)^{-1}  \sqrt{\frac {0.45\ GeV/cm^3}{\rho_{DM}}}
\eeq
for $m_a$ in a range $m_a=1.9-3.55$ $\mu$eV. Here $\rho_{DM}$ is the local dark matter density.
In our case we have  $|G_{a\gamma \gamma}|/(m_a/\mu eV)\simeq 1.4\times 10^{-16}$ (GeV$^{-1}$).
The upgrading of ADMX should be able to test  the axion parameters of the present scheme
\footnote{See ref.\cite{unificaxion}
and references therein for recent ideas of about axions in the context of fine-tuning.}.
This would be an important test of these ideas.

Let us finally comment that a possible problem for the axion in the local modulus $T$ to become a QCD axion
is moduli fixing. Indeed one may wonder whether the dynamics fixing the moduli could
also give a large mass to Im $T$. However this is not necessarily the case see  e.g.
\cite{Conlon:2006tq,Bobkov:2010rf,cicoli}.

\subsection{Cosmology}

The main new ingredient in this ISSB scheme is the large SUSY breaking scale 
$M_{SS}\simeq  5\times 10^{10}$ GeV and relatively low string scale $M_s\simeq 6\times 10^{14}$ GeV.
It is a true fact that having low-energy SUSY leads to a number of problems which are automatically solved
with such large SUSY scale. In particular 
there is no moduli, gravitino nor Polony problem.  

Another problem which is solved is the one first pointed out
in \cite{Kallosh:2004yh}.  This problem appears in string moduli fixing models like KKLT and other extensions
in which a supergravity scalar potential combined with other SUSY breaking effects (like antibranes) 
fix the moduli. Including the inflaton within such schemes leads to the conclusion that the Hubble scale at inflation
$H_I$ must verify $H_I<m_{3/2}$ in order for moduli fixing not to be destroyed. In a low scale SUSY model 
with $m_{3/2}<1$ TeV that poses a problem.  In our case however 
with $m_{3/2}\simeq 5\times 10^{10}$ GeV the problem disappears and inflation and KKLT type of
moduli fixing are easily compatible. From this point of view one could argue that inflation in
models with KKLT-type moduli fixing suggests  a SUSY scale $M_{SS}>10^{10}$ GeV.

\subsection{Neutrino masses}

Singlets playing the role of right-handed neutrinos may appear in  F-theory GUT schemes.
A natural source of masses for  right-handed neutrinos in this framework is 
string instantons see ref.\cite{stringinstantons,instantonrev}. 
In our case the instanton suppression is typically of order
$exp(-2\pi/g_s)$ which for $g_s=0.28$ may be of order $10^{-10}$. Thus one can generate right-handed 
neutrino masses of order 
 $ e^{-2\pi/g_s}\times M_s \ \simeq \ 10^{-10} M_s\ \simeq \    \times 10^4\ GeV$
for $g_s=0.28$ and $M_s\simeq 10^ {14}$ GeV,  as suggested by the size of gauge threshold corrections. This would require 
small Dirac neutrino masses to be compatible with experiment. On the other hand 
if the theory above $M_{SS}$ is an R-parity violating version of the MSSM, there may be additional sources of neutrino masses. 
In particular if there are R-parity violating terms of the form $M_{SS}(L_iH_u)$ the leptons remaining at low-energies
mix with the Higgsinos. After the Higgs gets a vev this induces see-saw Majorana neutrino masses of
order $M_W^2/M_{SS}$ which may also be consistent with the observed neutrino masses.

\section{High Scale SUSY versus Split SUSY}

\begin{figure}[t]
\begin{center}
\includegraphics[width=11cm]{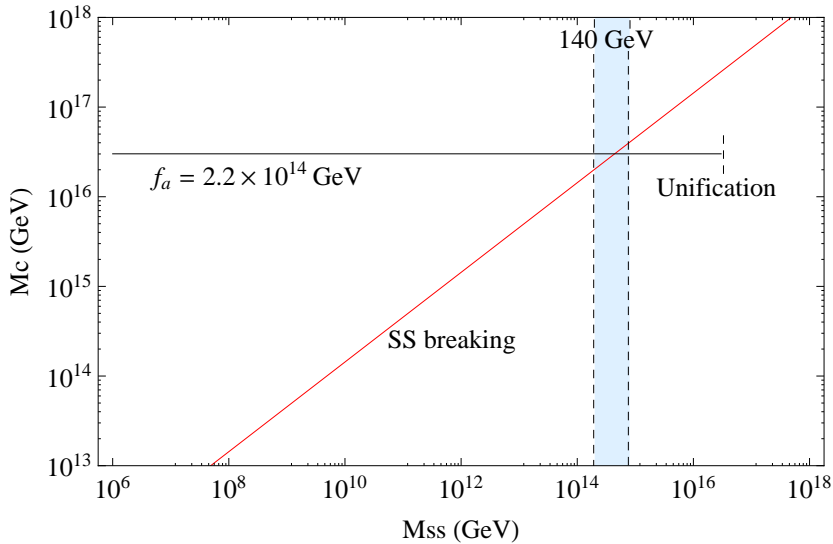}
\end{center}
\caption{Unification and SUSY breaking scales in Split-SUSY versus the Higgs mass. Flux SUSY breaking naturally 
prefers $M_{SS}\simeq 10^ {14}$ GeV, corresponding to a Higgs mass around 140 GeV.}
\label{splitplot}
\end{figure}

In previous sections we have concentrated in the ISSB case in which the theory below the 
SUSY breaking scale $M_{SS}$ is just the SM.  A different alternative in the literature is that of Split-SUSY 
\cite{splitsusy} in which one assumes that below $M_{SS}$ there  is the SM plus in addition 
the gauginos and Higgsinos.  The latter then get  SUSY-breaking masses in a region close to the
EW scale.
One can also repeat the analysis in this case.
Concerning gauge coupling unification, we just have to replace the $\beta$-function coefficients 
of the SM by $(b_1^{sp},b_2^{sp},b_3^{sp})=(45/6,-7/6,-5)$.  One then finds from eq.(\ref{rgeunif})
\beq
 12\ log \frac {M_{SS}}{M_{EW}} \ +\ 12 \  log \frac {M_{c}}{M_{SS}} \ =\ 2\pi \  \left(\frac {1}{\alpha_1(M_{EW})} 
- \frac {1}{\alpha_2(M_{EW})}  - \frac {2}{3\alpha_3(M_{EW})} \right) ,
\label{lineaunifsplit}
\eeq
i.e., the dependence on $M_{SS}$ cancels out. This means one can always choose  threshold effects
(i.e. value of $g_s$) such that one-loop unification occurs. One finds the unification scale is fixed at
$M_c\simeq 3\times 10^{16}$  GeV, for any $M_{SS}$ (see fig.\ref{splitplot}). One also obtains  $g_s=1-5$
as one goes from $M_{SS}=10^{14}$ GeV to $M_{SS}=1$ TeV. This means that the threshold corrections 
are in general small, as expected from the fact that the unification of the MSSM and Split-SUSY work
numerically in quite a similar way.
 
 As found e.g. in \cite{Giudice:2011cg} if the Higgs mass is in the range $124-126$ GeV, the value of
 $M_{SS}$ for Split-SUSY  is in the region $10^7-10^4$ GeV as one goes from tan$\beta=1$ to tan$\beta=50$.
 The renormalization of  tan$\beta$ above $M_{SS}$ 
 works exactly as in section 5. Repeating the analysis we find that tan$\beta$ remains close to one for energies above
 $10^7$ GeV. Below those energies the loop corrections become increasingly important and tan$\beta$ grows 
 sharply close to   100 TeV.  Although a detailed analysis would be required, we expect that a Higgs mass
 in the range 124-126 GeV will be compatible with the boundary condition eq.(\ref{lambdasusy}) for
 $M_{SS}\simeq 100$ TeV.  Thus,  if the present Higgs hints are confirmed, Split-SUSY would be equivalent
 for all practical purposes to  a fine-tuned MSSM with the scalar sparticles heavier than the SUSY fermions.

Note that , unlike the case of High Scale SUSY, 
 such low values for $M_{SS}$ are not generic if SUSY-breaking is induced by closed string 
 fluxes which yield in our case  from eq.(\ref{genericmss}) $M_{SS}\simeq 10^{14}$ GeV (see fig.\ref{splitplot}).
 With such high  value of $M_{SS}$ one expects from \cite{Giudice:2011cg} a Higgs mass around 140 GeV.
 So in order to be in agreement with a Higgs mass in the region 124-126 GeV the SUSY breaking flux
 effects should be substantially diluted. In this respect  Split-SUSY is not  worse than the
 MSSM which also requires flux suppression to get soft terms of order $M_{SS}\simeq 1$ TeV.   Concerning the axion decay constant, since in this case $M_c\simeq 10^{16}$ GeV one obtains 
 a high value $F_a\simeq 10^{14}$ difficult to reconcile with cosmological limits.
 On the other hand the theory is automatically safe against too fast proton decay through dimension 6 operators 
 due to the large value of
 the unification scale $M_c$ and Higgsinos/Neutralinos may provide for the dark matter.
 A summary of the scales in Split-SUSY is shown in fig. \ref{splitplot}.

%

A relevant question is whether one can obtain a Split-SUSY type of spectrum below $M_{SS}$
in string compactifications. 
We need a first stage of SUSY-breaking in which only the scalars (but the SM Higgs) get a mass
of order $M_{SS}$.  One simple way to get such masses is through open string fluxes
\cite{ignatios}, very much as in the example in appendix A. 
 A second stage of SUSY-breaking, perhaps from (suppressed) closed string fluxes could give rise
 to gaugino and Higgsino masses close to 1 TeV.

 As a summary, High scale SUSY breaking \cite{Hall:2009nd}  with $M_{SS}\simeq 10^{11}$ GeV looks like 
 more natural in the sense that this  value of $M_{SS}$ is consistent both with the Higgs mass
 hints {\it and} a generic value of SUSY-breaking effects from fluxes. Furthermore it is simpler 
 in that only one-step of SUSY breaking is needed
  \footnote{Note also  that 
 in Split-SUSY with a scale $M_{SS}\simeq 10^{5}-10^7$ GeV, within a $SU(5)$ scheme 
 one needs to have some doublet-triplet splitting mechanism. 
 Furthermore some  discrete symmetry like R-parity should also exist to forbid 
  fast proton decay, unless the R-parity violating couplings are extremely  suppressed.
 On the other hand in High Scale SUSY with $M_{SS}\simeq 10^{11}$ GeV the existence of doublet-triplet  splitting
 or R-parity conservation are  not necessary. This is particularly relevant since getting $SU(5)$ 
 vacua with doublet-triplet splitting and
 R-parity  conservation turns out to be non-trivial in string compactifications.}.
  If the Higgs hints are correct, Split-SUSY 
 becomes really a fine-tuned version of the MSSM.

\section{Final conclusions  and outlook}

The LHC results are already
giving us substantial information on the 
physics at the electro-weak scale. No new physics has been observed as yet
and, in particular, SUSY colored particles remain elusive. As the energy and luminosity is increased 
the experimental constraints imply quite strong conditions on the MSSM parameters which
need to be fine-tuned at the per-mil level. If the hints of a Higgs mass around 125 GeV are 
confirmed this fine-tuning will be strengthened. Such a Higgs mass  is a  bit too heavy for the MSSM and 
a bit too low for the SM high energy stability. 

In view that some level of fine-tuning is required a natural  question is whether after all 
one should accept a fine-tuning explanation for the hierarchy problem. 
One could argue that 
the size of the EW scale could be selected on anthropic grounds, in analogy to
Weinberg's anthropic arguments about the cosmological constant \cite{weinberg}.
It is certainly premature to give up on the SUSY solution of the hierarchy problem
but it makes sense to explore what the alternatives are.

While the fine-tuning  idea is worth exploring,  it does not imply to give up the supersymmetry idea at some level.
We know that supersymmetry is a basic ingredient in string theory 
and, even if it does not survive at the EW scale, it could have some role at higher 
energies. In particular in string compactifications non-SUSY theories have generically
tachyons in their spectrum and are hence unstable.  So it makes sense to investigate 
whether the theory becomes supersymmetric above some scale $M_{SS}$  
in between $M_{EW}$ and the string scale $M_s$. Guided by the apparent unification
of coupling constants we  would also like to know whether this could be combined with
some form of unification like $SU(5)$.

In this paper we  address  this issue within the context of string theory. In particular
we study the possible structure of mass scales in F-theory unified schemes in which a
version of $SU(5)$ unification is possible. We find that there are several pieces of information
pointing to a SUSY  structure above an intermediate scale $M_{SS}\simeq 5\times 10^{10}$ GeV
and a unification scale $M_c\simeq 3\times 10^{14}$ GeV.  F-theory $SU(5)$ unification differs 
from field theory unification in some important aspects.  The breaking of $SU(5)$ to the SM via
hypercharge fluxes generates very specific corrections  to the gauge couplings. Combining these corrections with the idea
of closed string flux SUSY breaking fixes the scale of SUSY breaking at that
intermediate scale. Interestingly, the extrapolation up in energies of the SM Higgs self-coupling $\lambda$ 
seems to vanish also around that scale, giving additional
evidence. This is also consistent with  the MSSM-like boundary 
conditions $\lambda=\frac {1}{8}(g_1^2+g_2^2)cos^22\beta$ if  tan$\beta \simeq 1$ at 
$M_{SS}$. We have shown how such value for tan$\beta = 1$ is expected in known sources 
of SUSY breaking  in string theory and is also stable under loop corrections. An additional support for an ISSB 
structure of scales
is obtained from the existence of a natural axion candidate with a decay constant
$F_a\simeq  M_c/(4\pi)^2\simeq 2\times 10^{12}$ GeV, which could solve the strong CP 
problem and constitute the observed dark matter.

The unification scale $M_c\simeq 3\times 10^{14}$ GeV is relatively low and in field theory 
$SU(5)$ would be ruled due to too fast proton decay via dimension 6 operators.
We have shown however that in the context of F-theory $SU(5)$ there is an effect
due to the fact that $SU(5)$ is broken by hypercharge fluxes, not a Higgs mechanism. 
In this case the  hypercharge fluxes modify  the profile wave functions of matter 
and $X,Y$ fields  suppressing the couplings which could mediate fast proton decay.
If the suppression is not excessive, proton decay e.g. through the standard
{\it non-SUSY} channel $p\rightarrow \pi^0 e^+$ could be measured in future experiments.
In the present context doublet-triplet splitting is not unavoidable, the  Higgs triplets
may remain massless below $M_c$ and get masses of order $M_{SS}\simeq 10^{11}$ 
GeV. They can give additional contributions to proton decay with a dominant
$p\rightarrow \mu^+K^0$ channel.

If this ISSB framework is correct, no new particles beyond those of the SM and the Higgs particle 
would be uncovered at  the LHC.  On the other hand, since the value of $M_{SS}$ within this
scheme is fixed by gauge coupling unification + closed string flux breaking, one may consider 
the vanishing of  the Higgs self-coupling at $M_{SS}\simeq 10^{11}$ GeV as a prediction of this framework.
At present the uncertainties on the Higgs and top quark masses do not allow for a definite
conclusion. On the other hand
more precise measurements of these quantities could confirm or exclude
these ideas. Since in the present scheme dark matter is provided by axions, no WIMP's are
required at the EW scale,  but  microwave cavity experiments like the  upgraded AMDX experiment,
could detect axions in the range here predicted. The observation of proton decay  could also provide additional support
for this scheme.

 \vspace{1.0cm}

\bigskip

\centerline{\bf \large Acknowledgments}

\vspace{.25cm}

We thank P.G.~C\'amara, A.~Font, and  A.~Uranga  for discussions. We also thank Daniela Herschmann for pointing out an error in eq.(\ref{calculogs}) of the previous version.
This work has been partially supported by the grants FPA 2009-09017, FPA 2009-07908, Consolider-CPAN (CSD2007-00042) from the MICINN, HEPHACOS-S2009/ESP1473 from the C.A. de Madrid and the contract ``UNILHC" PITN-GA-2009-237920 of the European Commission. F.M. is supported by the Ram\'on y Cajal programme through the grant RYC-2009-05096 and by the People Programme of FP7 (Marie Curie Auction) through the REA grant agreement PCIG10-GA-2011-304023. D.R. is supported through the FPU grant AP2010-5687.

\newpage

\appendix

\section{ Non SUSY  SM Type II orientifolds with open string fluxes}

One of the simplest constructions in string theory  yielding a SM group with three
quark lepton generations is that of toroidal Type IIA orientifolds. In these constructions
(see e.g.(\cite{interev,BOOK}) and references therein) the SM particle are localized at the intersections 
of D6-branes. In the simplest schemes one has   $3+2+1+1$ sets of D6-branes 
corresponding to a gauge group $U(3)\times U(2)\times U(1)\times U(1)$. The extra $U(1)$'s beyond hypercharge 
get massive through a Green-Schwarz mechanism and one is left with the gauge group of the standard model.
The worldvolume of D6-branes is $4+3$ dimensional with the three extra dimensions being wrapped as
1-cycles in each of the three tori in the $T^2\times T^2\times T^2$ extra  dimensions. The 
number of times the D6-brane wraps  the $(x^i, y^i)$, $i=1,2,3$  cycles in each of the three 
tori is parametrized by integer numbers 
$(n_1,m_1)\times (n_2,m_2)\times (n_3,m_3)$.  Chiral fermions and scalars appear at the brane
intersections and the multiplicity of generations appear because D6-branes intersect typically a multiple
number of times.
The number of intersections  between two branes $a$ and $b$ is given by the topological invariant
\beq
I_{ab} \ =\ (n_1^am_1^b-m_1^an_1^b)(n_2^am_2^b-m_2^an_2^b)(n_3^am_3^b-m_3^an_3^b)
\label{intersecciones}
\eeq
The intersection angle between two branes $a,b$  in the i-th torus is given 
by $tan^{-1}(m_i^aU_i/n_i^a)-tan^{-1}(m_i^bU_i/n_i^b)$, where $U_i=R_i^y/R_i^x$ 
are the three complex structure parameters of the tori.
One can easily find choices of brane wrapping numbers such that the obtained chiral fermions 
correspond to those of the SM with three generations, see \cite{Ibanez:2001nd}.
In particular there is a large family of  3-generation models of this class which may be obtained 
from the wrapping numbers in table \ref{solution}.
\begin{table}[htb] \footnotesize
\renewcommand{\arraystretch}{2.5}
\begin{center}
\begin{tabular}{|c||c|c|c|}
\hline
 $N_i$    &  $(n_i^1,m_i^1)$  &  $(n_i^2,m_i^2)$   & $(n_i^3,m_i^3)$ \\
\hline\hline $N_a=3$ & $(1/\beta ^1,0)$  &  $(n_a^2,\epsilon \beta^2)$ &
 $(1/\rho ,  1/2)$  \\
\hline $N_b=2$ &   $(n_b^1,-\epsilon \beta^1)$    &  $ (1/\beta^2,0)$  &  
$(1,3\rho /2)$   \\
\hline $N_c=1$ & $(n_c^1,3\rho \epsilon \beta^1)$  & 
 $(1/\beta^2,0)$  & $(0,1)$  \\
\hline $N_d=1$ &   $(1/\beta^1,0)$    &  $(n_d^2,-\beta^2\epsilon/\rho )$  &  
$(1, 3\rho /2)$   \\
\hline \end{tabular}
\end{center} \caption{ D6-brane wrapping numbers giving rise to a SM spectrum.
\label{solution} }
\end{table}
Here $N_i$ give the number of parallel branes giving rise to a gauge group $U(N_i)$.
Along with each stack of branes there should be additional {\it mirror } D6-branes 
in order for the brane configuration to be invariant under the orientifold operation.
Those mirror branes are obtained by flipping the sign of the $m$ component 
of the wrapping numbers and are not displayed in the table.
The general solutions 
 are parametrized by a phase $\epsilon =\pm1$, the NS background
on the first two tori $\beta^i=1-b^i=1,1/2$, four integers
$n_a^2,n_b^1,n_c^1,n_d^2$ and a parameter $\rho=1,1/3$.
The reader may check that e.g. there are three right-handed leptons  at the
intersection of branes c and d, i.e., $I_{cd}=3$ from equation (\ref{intersecciones}).
We are interested here in the Higgs sector.
The Higgs doublets appear from open strings exchanged between  the 
$b$ and $c$ branes (or the mirror brane $c^*$). As one can see in table \ref{solution} 
the $U(2)$ branes ($b,b^*$) are parallel 
to the ($c,c^*$) branes along  the second torus and hence they do not intersect.
However there are open strings which stretch in between  both sets of branes
and which lead to  light scalars when the distance $Z_2$ in the second torus
is small. In particular there are Higgs doublets $h_u,h_d$ from $(bc)$ intersections 
and $H_u,H_d$  from $(bc^*)$ intersections, although for some choices of wrapping numbers only 
the $h$'s or the $H$'s survive.
In particular there are the scalar states 
{\small \beqa
\begin{array}{cc}
{\rm \bf State} \quad & \quad {\bf Mass^2} \\
(-1+\vartheta_1, 0, \vartheta_3, 0) & \alpha' {\rm (Mass)}^2 =
  { {Z_2}\over {4\pi ^2}}\ +\ \frac{1}{2}(\vartheta_3 - \vartheta_1) \\
(\vartheta_1, 0, -1+ \vartheta_3, 0) & \alpha' {\rm (Mass)}^2 =
  { {Z_2}\over {4\pi ^2 }}\ +\ \frac{1}{2}(\vartheta_1 - \vartheta_3) \\
\label{Higgsmasses}
\end{array}
\eeqa}
where $Z_2$ is the distance$^2$ (in $\alpha '$ units) 
 in transverse space along the second torus.
$\vartheta_1 $ and $\vartheta_3$ are the relative angles between the
$b$ and $c$ (or $b$ and $c^*$) in the first and third complex planes.
The states are defined above as vectors in the $SO(8)$ light-cone target
space of Type IIA string theory \cite{BOOK}. There are fermionic states also of the form
{\small \beqa
\begin{array}{cc}
{\rm \bf State} \quad & \quad {\bf Mass^2} \\
(-1/2+\vartheta_1, \mp 1/2 , -1/2+\vartheta_3, \pm 1/2 ) &  {\rm (Mass)}^2 =
  { {Z_2}\over {4\pi ^2\alpha '}}\  \\
\label{Higgsinomasses} 
\end{array}
\eeqa}
This Higgs system may be understood as  
massive $\cn=2$ Hypermultiplets containing respectively the $h_i$ and/or $H_i$ 
scalars along with the above fermions. The above scalar spectrum corresponds
to the following mass terms in the effective potential:
\beqa
V_2\ =\ m_H^2 (|H_1|^2+|H_2|^2)\ +\ m_h^2 (|h_1|^2+|h_2|^2)\ + \nonumber \\
+\ m_{3H}^2 H_1H_2+h.c. \ +\ m_{3h}^2h_1h_2+h.c.
\label{Higgspot}
\eeqa
where:
\beqa
 {m_h}^2 \ =\ { {Z_2^{(bc)}}\over {4\pi ^2\alpha '}}\ & ; & \ 
{m_H}^2 \ =\ {{Z_2^{(bc^*)}}\over {4\pi ^2\alpha '}}\nonumber \\
m_{3h}^2\ =\ \frac{1}{2\alpha '}|\vartheta_1^{(bc)}-\vartheta_3^{(bc)}| \ & ;&
m_{3H}^2\ =\ \frac{1}{2\alpha '}|\vartheta_1^{(bc^*)}-\vartheta_3^{(bc^*)}|
\label{masillas}
\eeqa
Notice that each of the two Higgs systems 
have a quadratic potential similar to that of the  
MSSM.
In fact one also expects the quartic potential to be identical to that
of the MSSM.
In our case the mass parameters of the potential have an
interesting  geometrical  interpretation in terms of the brane distances 
and intersection angles.  In the main text we consider for simplicity the presence of just one set $h$ of
Higgs multiplets. This may be achieved by appropriate choice of the wrapping numbers.

\section{RGE solutions}

Here we display the definition of the functions appearing
in the solution of the RGE in ref.\cite{ilm}. 
We define 
\beq
E(t)\ =\ (1+\beta_3t)^{16/(3b_3)}(1+\beta_2t)^{3/(b_2)}(1+\beta_1t)^{13/(9b_1)}
\ \ ,\ \ 
F(t)=\int_0^t E(t')dt'
\eeq
with $\beta_i=\alpha_i(0)b_i/(4\pi)$
and $t=2\log(M_c/M_{SS})$. 
Here we have $(b_1,b_2,b_3)=(11,1,-3)$.  We also define $Y_t=h_t^2/(4\pi)^2$
with $Y_0=Y_t(0)$ and $\alpha_0=\alpha_i(0)=\alpha(0)$ for $i=2,3$,
 $\alpha_1(0)=(3/5)\alpha(0)$.
Here $\alpha_0$ is the unified coupling at $M_c$.
In our case the couplings do not strictly unify, only up to $5\%$ corrections.
In the numerical computations we take the average value of the 
three couplings at $M_c$, which is enough for our purposes.
We then define the functions in eqs.(\ref{ilm1},\ref{ilm2})
\beqa
q(t)^2  & =&  \frac {1}{(1+6Y_0F(t))^{1/2}}(1+\beta_2t)^{3/b_2}
(1+\beta_1t)^{1/b_1} \ ;\ 
h(t) = \frac {1}{2}(3/D(t)-1)  \nonumber \\ 
k(t)\ & = & \ \frac {3Y_0F(t)}{D(t)^2} \ ;\
f(t)\ =\ -\frac {6Y_0H_3(t)}{D(t)^2} \ ;\ D(t)\ =\ (1+6Y_0F(t)) \\ \nonumber
e(t)\ & =& \ \frac {3}{2} \left( \frac {(G_1(t)+Y_0G_2(t))}{D(t)} 
\ +\ \frac {(H_2(t)+6Y_0H_4(t))^2}{3D(t)^2} \ +\ H_8\right) \nonumber \\
\eeqa
and  the functions $g,H_2,H_3,H_4,G_1,G_2$ and $H_8$ are {\it independent of
the top Yukawa}  coupling, only depend on the gauge coupling constants and are given by
\beqa
g(t) \  &=&\ \frac {3}{2} \frac {\alpha_2(0)}{4\pi} f_2(t) \  + \ \frac {1}{2} \frac {\alpha_1(0)}{4\pi}
f_1(t) \nonumber \\
H_2(t) \ & =& \  \frac {\alpha_0}{4\pi}\left(\frac {16}{3}h_3(t)\ +\ 3h_2(t)\ +\ \frac {13}{15}h_1(t) \right) \nonumber \\
H_3(t) \  &=& \ tE(t)\ -\ F(t) \nonumber \\
H_4(t)\ &=& \ F(t)H_2(t) \ -\ H_3(t) \nonumber \\
H_5(t) \  & = & \ \frac {\alpha_0}{4\pi} \left(-\frac{16}{3}f_3(t) \ +\ 6 f_2(t) \ -\ \frac {22}{15} f_1(t) 
\right) \nonumber \\
H_6(t) \ & =& \ \int_0^t\ H_2(t')^2 E(t')dt' \nonumber \\
H_8(t) \ & = & \ \frac {\alpha_0}{4\pi }\ \left( -\frac {8}{3}f_3(t)\ +\ f_2(t) \ -\ \frac {1}{3} f_1(t) \right) \nonumber \\
G_1(t) \ & = &\ F_2(t) \ -\ \frac {1}{3} H_2(t)^2 \nonumber \\
G_2(t) \ & = & \ 6F_3(t) \ -\ F_4(t) \ -\ 4H_2(t)H_4(t) \ +\ 2F(t)H_2(t)^2 \ 
\ -\ 2H_6(t) \nonumber \\
F_2(t) \ & = & \ \frac {\alpha_0}{4\pi } \ \left( \frac {8}{3}f_3(t) \ +\ \frac {8}{15}f_1(t) \right)
\nonumber \\
F_3(t) \ & = &\ F(t)F_2(t)\ -\ \int_0^t E(t')F_2(t') dt' \nonumber \\
F_4(t) \ & = & \ \int_0^t E(t')H_5(t') dt' 
  \eeqa
  where $f_i(t)$ and $h_i(t)$ are defined by
  \beq 
  f_i(t) \ =\ \frac {1}{\beta_i}(1\ -\ \frac {1}{(1+\beta_it)^2}) 
  \ ;\ 
  h_i(t) \ =\ \frac {t}{(1+\beta_it)} \ .
  \eeq
   The low energy of the top mass may be obtained from 
   the solutions of the one-loop renormalization group equations,
   divided into two pieces, SUSY and non-SUSY,  i.e. (here $Y_t=h_t^2/(16\pi^2)$)
\beq
Y_t(m_t) \ =\ sin^2\beta Y_t(M_{SS})  \frac  {E'(t_{EW})}{(1+(9/2) sin^2\beta Y_t(M_{SS})F'(t_{EW}))}
\eeq
where
\beq
Y_t(M_{SS}) \ =\   Y_t(M_c)  \frac  {E(t_{SS})}{(1+6  Y_t(M_c)F(t_{SS}))} 
\label{topyukis}
  \eeq
where  the functions $E,F$ are as defined above, with $t_{SS}=2log(M_c/M_{SS})$
and $t_{EW}=2log(M_{SS}/M_{EW})$.
The functions $E',F'$ are analogous to $E,F$ but replacing the $b_i$ and anomalous dimensions
by the non-SUSY ones, i.e. 
\beq
E'(t)\ =\ (1+\beta'_3t)^{8/(b^{NS}_3)}(1+\beta'_2t)^{9/(4b^{NS}_2)}(1+\beta'_1t)^{17/(12b^{NS}_1)}
\ \ ,\ \ 
F'(t)=\int_0^t E'(t')dt'
\eeq
with $\beta'_i=\alpha_i(M_{SS})b^{NS}_i/(4\pi)$
and $t=t_{EW}$. 
We have now $b_i^{NS}=(41/6,-19/6,-7)$.
For the anomalous dimensions we have made the change in the definition of $E(t)$ 
$(13/9,3,16/3)$ $\rightarrow(17/12,9/4,8)$.
Then $m_t(m_t) = h_t(m_t) <H>= h_t(m_t)(174.1)$ GeV. For this particular computation we take actually 
$t_{EW}=2log(M_{SS}/(173.2 GeV))$.

In the case of Split-SUSY the same formulae in eq.(\ref{topyukis}) apply replacing $E',F'$ by $E'',F''$ given by
\beq
E''(t)\ =\ (1+\beta''_3t)^{8/(b^{sp}_3)}(1+\beta''_2t)^{9/(4b^{sp}_2)}(1+\beta''_1t)^{17/(12b^{sp}_1)}
\ \ ,\ \ 
F''(t)=\int_0^t E''(t')dt'
\eeq
with $\beta''_i=\alpha_i(M_{SS})b^{sp}_i/(4\pi)$
and $t=t_{EW}$, 
where  now $b_i^{sp}=(45/6,-19/6,-5)$.

\section{Wave functions and proton decay}
\label{ap:proton}
  
 In F-theory GUT models most couplings of the 4d effective theory are obtained by dimensional reduction over the GUT 4-cycle $S$. In particular, to compute particle interactions one needs to consider the internal wavefunction of the 4d fields of the theory, which typically have a non-trivial profile over $S$, and compute overlaps of these wavefunction such as (\ref{yukawas}) or (\ref{d3coups}).
 
The internal wavefunction profile of the 4d particles is found by solving the corresponding equations of motion, which in turn arise from the 8d 7-brane action found in \cite{BHV}. For 4d massless fermions, one can express such internal equations of motion as a Dirac like equation, namely as \cite{afim}
\beq
{\bf D_A} \Psi\, =\, 0
\label{Dirac9}
\eeq
with
\beq
{\bf D_A}\, =\, 
\left(
\begin{array}{cccc}
0 & D_x & D_y & i \phi^* \\
-D_x & 0 & i \phi & D_{\bar{y}} \\
-D_y & -i \phi^* & 0 & -D_{\bar{x}} \\
-i \phi & -D_{\bar{y}} & D_{\bar{x}} & 0
\end{array}
\right)
\quad \quad
\Psi\ \, \equiv\, 
\left(
\begin{array}{c}
- \sqrt{2}\, \eta \\ \psi_{\bar{x}} \\ \psi_{\bar{y}} \\ \chi_{xy}
\end{array}
\right)\label{matrixDirac}
\eeq
where $D  = \p - i\langle A \rangle$ is the covariant derivative of the 7-brane 8d gauge theory. The components of $D$ that appear in (\ref{matrixDirac}) are along complex coordinates $(x,y)$ of the internal 4-cycle $S$ where our GUT 7-brane is wrapped. Hence, we have that $\langle A_{x,y} \rangle$ is non-zero because there are internal worldvolume fluxes $F = dA$ threading such 4-cycle, and it is precisely in this way how the hypercharge flux $F_Y$ as well as other fluxes enter the equations of motion. If the fermion arises from a matter curve $\Sigma_\a$ then $\phi = m^2 f_\a(x,y)$, where $f_\a$  stands for a holomorphic function such that $\Sigma_\a = \{f_\a(x,y)=0\}$, and $m$ is a mass scale (the intersection slope) of the order of $m_*$. If the fermions are in the bulk of $S$ and so not related to any matter curve then $\phi =0$. 

Each of the components of the vector $\Psi$ contains the degrees of freedom of a 4d chiral fermion, which is related to the fact that these equations of motion arise from fermions in higher dimensions. Typically, 4d chiral fermions have a vanishing component $\eta$ and the other three components non-vanishing, while the opposite happens for 4d gauginos. 

Let us for instance consider the case where fermionic zero modes in the representation {\bf 10} of SU(5) arise from a matter curve given by $\Sigma_{10} = \{y=0\}$. Let us also consider a worldvolume flux of the form\footnote{See \cite{eterno} for more details on this F-theory model.} 
\begin{eqnarray}
\label{totalflux}
\langle F \rangle& = & i(dy\wedge d\bar y - dx\wedge d\bar x) Q_P + i(dx\wedge d\bar y+dy\wedge d\bar x)Q_S \\ \nonumber
 & & -\,  i(dy\wedge d\bar y + dx\wedge d\bar x) M_{xy} Q_{10}
\label{totalflux}
\end{eqnarray}
where $Q_{10}$ is a gauge generator such that fermions at $\Sigma_{10}$ have charge $+1$, and we have defined 
\begin{eqnarray}
Q_P & = & -M Q_{10} + \tilde{N}_Y Q_Y \hspace*{1cm} M \, = \, \frac{1}{2}(M_y-M_x)  \\
\label{qpdef}
Q_S & = & N_FQ_{10}+N_YQ_Y \hspace*{1.1cm} M_{xy}\, = \, \frac{1}{2}(M_x+M_y)
\label{qsdef}
\end{eqnarray}
with $Q_Y$ the hypercharge generator and $M_x < 0 < M_y$. Here $M$, $M_{xy}$, $N_F$, $N_Y$ and $\tilde N_Y$ are flux densities, which in a local patch around $x=y=0$ one can approximate to be constant. One can then check that a zero mode solution for (\ref{Dirac9}) is given by \cite{cdp,eterno}
\beq
\Psi_{10}^i\,=\,
\left(
\begin{array}{c}
0 \\ \vec{v}
\end{array}
\right)\,
\psi^i_{10},
\quad 
\psi^i_{10} \, =\, \gamma_{10}^i m_*^{4-i} \, (x + \zeta y)^{3-i}  \, e^{\frac{M_{x}+q_Y\tilde{N}_Y}{2}|x|^2} e^{\frac{M_{y}-q_Y\tilde{N}_Y}{2}|y|^2} e^{\lam y(\bar{y} - \zeta\bar{x})}  e^{- q_S {\rm Re} (x\bar y)}
\label{wave10ap}
\eeq
where $q_Y$ is the hypercharge of each particle inside the 10-plet and $q_S= N_F + q_Y N_Y$ is (\ref{qsdef}) evaluated for each of them. In addition  $\lam$ is the lowest (negative) eigenvalue of the matrix
\begin{equation}
\left( \begin{array}{ccc}
 - M_{x} - q_Y\tilde{N}_Y & q_S & 0 \\
 q_{S} & - M_{y}+q_Y\tilde{N}_Y & im^2 \\
 0 & -im^2 & 0 
 \end{array} \right)
\end{equation}
$\vec{v}$ is the corresponding unit eigenvector and $\zeta = - q_S/(\lam+M_{x}+q_Y\tilde{N}_Y)$. Finally, $\gamma_{10}^i$ is a normalization factor such that
\beq
\langle \Psi^i_{10}|  \Psi^j_{10} \rangle\, =\, 2\,m_*^2\,||\vec v||^2 \int_S (\psi_{10}^i)^*\psi_{10}^j\, {\rm d vol}_S\, =\,\delta^{ij}
\eeq
and $i$ labels the wavefunction for the $i^{\rm th}$ family of 10-plets. 

As the flux (\ref{totalflux}) is quantized over each curve of $S$, taking the volume of $S$ reasonably large takes us to the regime $m^2 \gg |M|, |M_{xy}|, |N_F|, |N_Y|, |\tilde N_Y|$, as oftentimes assumed in F-theory models. In this limit we have $\lam  \raw - m^2$ and $\zeta \raw 0$, and so our wavefunction can be approximated by
\beq
\psi^i_{10} \, = \, \gamma_{10}^{i} \, m_*^{4-i} \, x^{3-i}  \ e^{\frac{M_{x}+q_Y\tilde{N}_Y}{2}|x|^2} e^{- m^2 |y|^2 - q_S {\rm Re} (x\bar y)}.
\label{wave10ap2}
\eeq
as done in the main text.

A similar analysis can be made to solve the wavefunction for the $X,Y$ bosons, which are eigenfunctions of the Laplacian
\beq
\Delta \Phi_{X,Y}\, =\, - \frac{5}{6} \mu \, \Phi_{X,Y} \quad \quad \Delta\, =\, \{D_x, D_{\bar{x}}\} + \{ D_y, D_{\bar{y}}\}
\eeq
with $\mu=\sqrt{N_Y^2+\tilde N_Y^2}$ and now the covariant derivatives are constructed from the piece of the vector potential $A$ proportional to $Q_Y$, as the $X,Y$ bosons are neutral under $Q_{10}$. A solution to the above equation is given by
\beq
\Phi_{X,Y} (x,y)  = \gamma_{X,Y}\, m_*\, e^{-\frac{5}{12} \mu \,(|x|^2+|y|^2)}
\eeq
where $\gamma_{X,Y}$ is again a normalization factor. 

Finally, let us compute the coupling $\Gamma_1^{ij}$ leading to the dimension 6 proton decay operator. As discussed in section \ref{proton}, we need to compute an overlap integral of the form
\beq
A_1 = 2\, m_*\, \vec v^{\dagger}\cdot \vec v  \int_S (\psi_{10}^{1})^* \psi_{10}^{1} \Phi_{X,Y} 
\label{A1}
\eeq
where for concreteness we have specified to the first family. The two wavefunctions $\Psi_{10}^1$ in (\ref{A1}) are of the form (\ref{wave10ap}) but because of gauge invariance they must correspond to particles of the 10-plet with different hypercharge. Hence we have $(q_S, q_P, \lam, \zeta) =  (q_{S_1}, q_{P_1}, \lam_1, \zeta_1)$ for one of them and $(q_S, q_P, \lam,\zeta) =  (q_{S_2}, q_{P_2}, \lam_2,\zeta_2)$ for the other one, where we have abbreviated $q_P = -M + q_Y\tilde{N}_Y$. Notice that this yields different vectors $\vec v$ for each particle. Instead of doing the $\delta$-function approximation as in the main text, let us compute the above integral with the original wavefunction (\ref{wave10ap}). We find that
\begin{eqnarray}
\hspace*{-1.5cm}\nonumber
A_1&=&\frac{20\sqrt 2 \pi^2 }{6}g_s \mu\, m_*^6\left ( 1+\frac{\lam_1\zeta_1\lam_2\zeta_2}{m^4}+\frac{\lam_1\lam_2}{m^4} \right )\left (\frac{M_{xy}+q_{P_1}}{m_*^2} \right )\left (\frac{M_{xy}+q_{P_2}}{m_*^2} \right )
\end{eqnarray}
\vspace*{-.75cm}
\begin{eqnarray}
\hspace*{-.5cm}\nonumber
\left ( \frac{2\lam_1(M_{xy}+q_{P_1})+M_{xy}^2-q_{P_1}^2-(M_{xy}+q_{P_1})^2\zeta_1^2}{m^4+\lam_1^2(1+\zeta_1^2)} \right )^{1/2}\left ( \frac{2\lam_2(M_{xy}+q_{P_2})+M_{xy}^2-q_{P_2}^2-(M_{xy}+q_{P_2})^2\zeta_2^2}{m^4+\lam_2^2(1+\zeta_2^2)} \right )^{1/2}
\end{eqnarray}
\vspace*{-.75cm}
\begin{eqnarray}
\hspace*{-0.5cm}\nonumber
\frac{\left (\frac{5\mu}{12}-M_{xy}+\frac{q_{P_1}+q_{P_2}}{2}-\lam_1-\lam_2-(\frac{q_{S_1}+q_{S_2}}{2}+\lam_1\zeta_1)\zeta_1-(\frac{q_{S_1}+q_{S_2}}{2}+\lam_2\zeta_2)\zeta_2+(\frac{5\mu}{12}-M_{xy}-\frac{q_{P_1}+q_{P_2}}{2})\zeta_1\zeta_2\right )^2}{\left ( (\frac{5\mu}{12}-M_{xy}-\frac{q_{P_1}+q_{P_2}}{2})(\frac{5\mu}{12}-M_{xy}+\frac{q_{P_1}+q_{P_2}}{2}-\lam_1-\lam_2)-(\frac{q_{S_1}+q_{S_2}}{2}+\lam_1\zeta_1)(\frac{q_{S_1}+q_{S_2}}{2}+\lam_2\zeta_2) \right )^3}
\end{eqnarray}
Taking the limit $M,M_{xy},N_F,N_Y,\tilde N_Y\ll m^2$, the leading term of this expression is 
\beq
|A_1| \approx \frac{5\mu}{6\sqrt 2 \pi m_*^2}\frac{(-M_{x}-q_{Y_1}\tilde{N}_Y)^{3/2}(-M_{x}-q_{Y_2}\tilde{N}_Y)^{3/2}}{(\frac{5\mu}{12}-M_{x}-\frac{q_{Y_1}+q_{Y_2}}{2}\tilde{N}_Y)^3}
\eeq
reproducing eq.(\ref{final}) for $i=1$.

\end{document}